\documentclass[11pt]{article}

\usepackage[a4paper,margin=1in]{geometry}

\usepackage[T1]{fontenc}
\usepackage{newtxtext,newtxmath}
\usepackage{microtype}

\usepackage{xcolor}

\usepackage{mathtools}
\usepackage{bm}
\usepackage{siunitx}
\usepackage{graphicx}
\usepackage{booktabs}
\usepackage{tabularx}
\usepackage[font=small,labelfont=bf]{caption}
\usepackage{subcaption}

\usepackage{authblk}

\usepackage[hidelinks]{hyperref}
\hypersetup{
  pdftitle   = {Quantum Magnetometers for Infrastructure Inspection and Monitoring},
  pdfauthor  = {Muhammad Mahmudul Hasan; Ingrid Torres; Alex Krasnok},
  pdfsubject = {Review},
  pdfkeywords= {quantum sensing, magnetometry, nondestructive evaluation, structural health monitoring}
}
\usepackage[nameinlink,noabbrev]{cleveref}
\title{\bfseries Quantum Magnetometers for Infrastructure Inspection and Monitoring}

\renewcommand\thefootnote{\fnsymbol{footnote}}
\author[1]{Muhammad Mahmudul Hasan}
\author[1]{Ingrid Torres}
\author[1]{Alex Krasnok\thanks{Corresponding author: \href{mailto:akrasnok@fiu.edu}{akrasnok@fiu.edu}.}}
\affil[1]{Department of Electrical Engineering, Florida International University, Miami, FL 33174, USA}
\date{}

\usepackage[hidelinks]{hyperref}
\hypersetup{
  pdftitle   = {Quantum Magnetometers for Infrastructure Inspection and Monitoring},
  pdfauthor  = {Muhammad Mahmudul Hasan; Ingrid Torres; Alex Krasnok},
  pdfsubject = {Review},
  pdfkeywords= {quantum sensing, magnetometry, nondestructive evaluation, structural health monitoring}
}
\usepackage[nameinlink,noabbrev]{cleveref}

\begin{document}
\maketitle

% Restore numeric footnotes after the title block (so later footnotes are normal)
\setcounter{footnote}{0}
\renewcommand\thefootnote{\arabic{footnote}}

% ---------------------------
% Abstract (revised)
% ---------------------------
\begin{abstract}
Damage in infrastructure is often hidden until it becomes costly or dangerous. Common examples include corrosion under insulation, early fatigue damage in steel, corrosion of embedded reinforcement, and abnormal current flow in batteries and power equipment. Magnetic methods are attractive because they can sense through coatings, insulation, and concrete cover without couplants, but field performance is often limited by lift-off, low-frequency drift, background magnetic noise, and the weak low-frequency response of pickup coils. This review examines two room-temperature quantum receiver platforms: optically pumped atomic magnetometers (OPMs) and nitrogen-vacancy (NV) diamond magnetometers. Rather than treating them as stand-alone sensors, we compare them as parts of a full measurement chain that includes source physics, geometry, readout, calibration, and interpretation. The literature is organized into four magnetic signal classes: driven induction responses, leakage fields in magnetic flux leakage inspection, passive self-fields linked to stress or corrosion, and fields produced by operational currents. OPMs are strongest for low-frequency, phase-referenced induction measurements, while NV sensors are strongest for near-surface field mapping, vector or gradient measurements, and differential current sensing in compact solid-state heads. Across all applications, deployment depends less on best-case sensitivity than on usable bandwidth, dynamic range, background rejection, geometry control, calibration, and validation. The clearest path to field use is therefore robust instrument engineering tied to qualification methods that reflect real inspection conditions.
\end{abstract}

% ---------------------------
% Introduction (revised)
% ---------------------------
\section{Introduction}
\label{sec:intro}

{ 
Hidden degradation is costly precisely because it often remains inaccessible to direct visual inspection until repair becomes expensive, service is disrupted, or safety margins are already reduced. Corrosion alone is estimated to cost on the order of trillions of dollars annually worldwide, and its burden spans pipelines, storage tanks, bridges, reinforced concrete, and electrical infrastructure \cite{NACEIMPACT2016}. In U.S. pipeline systems, corrosion has historically accounted for a substantial share of reported incidents, which is why early detection is not a marginal maintenance issue but a core risk-control problem \cite{PHMSA_CorrosionFactsheet}. Civil assets face the same pressure: ASCE's 2025 Report Card assigns an overall grade of C and emphasizes the need to move from reactive repair toward planned renewal \cite{ASCE2025ReportCard}. Magnetic methods are attractive in this setting because they can interrogate conductive and ferromagnetic assets through coatings, insulation, cladding, and concrete cover without couplants, but their value in the field depends on whether the magnetic signal of interest survives stand-off, drift, and environmental interference long enough to support a decision.
}

{ 
In this review, \emph{inspection} denotes an episodic measurement carried out to support a maintenance or repair decision at a particular time, whereas \emph{monitoring} denotes repeated measurements intended to detect change on the same asset over time. The distinction is operationally important. Inspection fails when weak signals cannot be separated from background fields, motion pickup, or uncontrolled geometry. Monitoring fails when an apparent trend is produced by drift in gain, lift-off, or alignment rather than by real material change. In both cases, repeatability under field variability is more important than the best laboratory noise floor, because maintenance decisions must be traceable to a stable procedure rather than to a single favorable scan \cite{abdollahi2025non}.
}

\begin{figure}[t]
  \centering
  \includegraphics[width=0.95\textwidth]{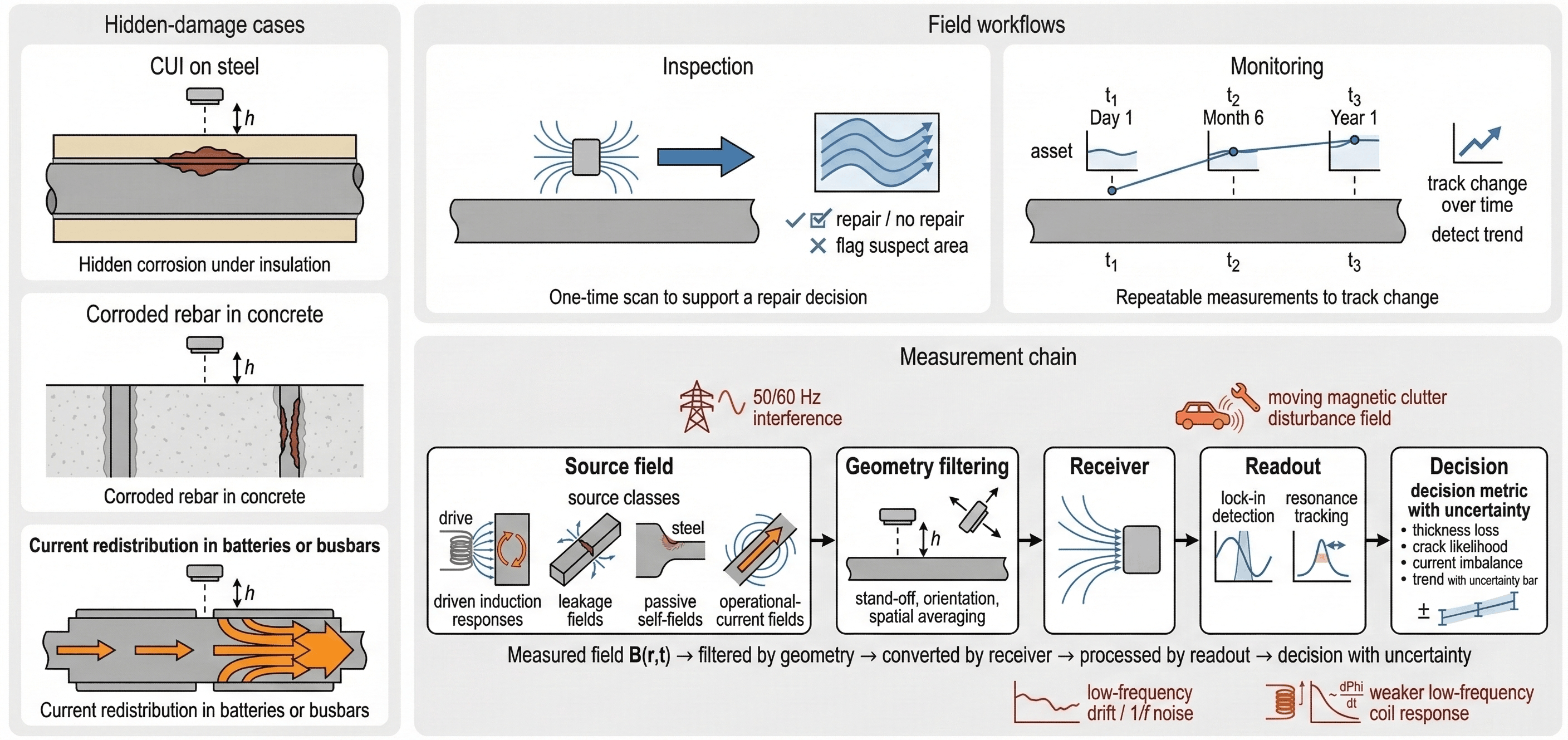}
  \caption{ {Infrastructure magnetometry overview. Left: representative hidden-damage cases that motivate sensing at stand-off $h$, including corrosion under insulation (CUI) on steel, corroded rebar in concrete, and current redistribution in batteries or busbars. Top: two field workflows---inspection (a one-time scan to support a repair decision) and monitoring (repeatable measurements to track change). Bottom: receiver-centric measurement chain and the four recurring source classes used throughout this review: driven induction responses, leakage fields, passive self-fields, and operational-current fields. The measured field $B(\mathbf{r},t)$ is filtered by stand-off and orientation, converted by the receiver, and processed by readout (lock-in detection or resonance tracking) into a decision metric with uncertainty. Key field constraints include 50/60~Hz interference and moving magnetic clutter, low-frequency drift/$1/f$ noise, and the reduced low-frequency response of inductive pickup ($\propto d\Phi/dt$).}}
  \label{fig:intro_overview}
\end{figure}

{ 
From a measurement standpoint, infrastructure magnetometry is not one problem but several recurring signal classes. Driven induction methods apply an alternating-current (AC) magnetic field and measure the secondary response from induced currents. Magnetic flux leakage (MFL) inspection measures stray fields caused by defects in a deliberately magnetized ferromagnet. Passive methods attempt to exploit residual, stress-linked, or corrosion-linked self-fields under ambient geomagnetic bias. Operational-current measurements infer current paths from the fields produced by the currents themselves. The same asset may support more than one class. A pipeline wall, for example, can be screened by induction, inspected by MFL, and monitored through passive magnetic signatures, but each mode imposes different receiver requirements and different calibration burdens.
}

{ 
Conventional magnetic and electromagnetic nondestructive evaluation (NDE) already provides effective baseline methods for many of these tasks. MFL is widely used for cracks and metal loss in ferromagnetic steel, but its signatures are strongly distorted by lift-off because defect-related fields decay rapidly with stand-off distance \cite{Feng2022MFLReview}. Induction and eddy-current methods are powerful because they can probe through coatings and nonconductive layers, yet their receiver chain becomes progressively more difficult at low frequency: a pickup coil is fundamentally a voltage receiver, with $V_{\mathrm{coil}}\propto -d\Phi/dt \sim \omega \Phi$, so reducing excitation frequency also reduces the directly induced signal. Many solid-state magnetic sensors further suffer from drift and low-frequency $1/f$ noise in the Hz--kHz band \cite{Wei2021FluxgateReview,Guedes2018HybridGMR}. In corrosion-under-insulation (CUI) workflows, pulsed eddy current testing (PECT) remains a practical benchmark because it can screen through insulation, but it still faces strong sensitivity to lift-off, layered geometry, and the usual trade-offs between penetration depth, spatial resolution, and inspection speed \cite{Sophian2017PECTReview,API_RP_583}.
}

{ 
A clearer way to compare sensing platforms is therefore to treat the full system as a measurement chain rather than to compare stand-alone sensor specifications. Source physics and excitation determine the magnetic field that actually exists in space and time. Geometry filters that field through stand-off, orientation, and spatial averaging. The receiver converts one component---or, in more advanced heads, a vector or gradient of the field---into an optical or electrical observable with finite noise, bandwidth, and dynamic range. Readout electronics then recover amplitude, phase, resonance frequency, or other derived observables, which must still be mapped into thickness loss, crack likelihood, tendon condition, or current redistribution through a forward model and a calibration procedure. In real assets, the dominant errors are often geometric (lift-off and orientation), environmental (time-varying background fields and nearby currents), or procedural (magnetization history and scan repeatability), not the advertised best-case noise floor of the sensing element itself.
}

{ 
Quantum magnetometry is not new to NDE. Superconducting quantum interference device (SQUID) receivers established several system ideas that remain central---gradiometry, background rejection, inversion from field maps, and the use of magnetic receivers in eddy-current or leakage-field inspection---but cryogenic operation limits broad infrastructure deployment \cite{Jenks1997SQUIDNDE}. The practical shift in the last decade is that high-sensitivity quantum receivers can now operate at room temperature and can be packaged into sensor heads that are plausible for scanning and field use.
}

{ 
This review focuses on the two room-temperature quantum receiver platforms that presently have the clearest path to infrastructure sensing: optically pumped atomic magnetometers (OPMs) and NV-center diamond magnetometers. Both can deliver strong low-frequency sensitivity and can be engineered for unshielded operation through differential readout, vector sensing, and gradiometry \cite{Tierney2019OPMReview,BudkerRomalis2007OpticalMagnetometry,Degen2017QuantumSensing,Barry2020Sensitivity}. They do not alter the source physics that set skin depth, leakage-field decay, or inverse-problem ill-posedness, but they can widen the usable operating window when conventional receivers become limited by low-frequency noise, drift, or bandwidth. The distinction between the two platforms is already visible at the systems level: OPMs are presently strongest where low-frequency, phase-referenced induction measurements dominate, whereas NV sensors are presently strongest where compact solid-state heads, small stand-off, vector information, or gradient information are decisive. In both cases, unshielded-operation engineering is part of the sensing method, because measurement quality in realistic environments is often limited more by magnetic backgrounds, vibration, nearby currents, and motion-induced pickup than by intrinsic sensor noise \cite{Fu2020ChallengingEnvironments}.
}

A second translation gap is interpretation and validation. In maintenance practice, a field map is rarely the desired output. Decisions are typically framed in terms of thickness loss, crack probability, tendon condition, rebar corrosion state, or current redistribution. Those outputs require calibration, a geometry-aware forward model, and an uncertainty statement. New receiver platforms must therefore be evaluated with acceptance pathways that resemble those used for other NDE systems, including documented procedures, representative artifacts, and probability-of-detection (POD) style metrics when appropriate \cite{MIL_HDBK_1823_ASSIST,ASTM_E2862_23}. This is one reason to organize the literature by \emph{signal class} rather than by sensor novelty: the available signal determines the task band, dynamic range, background rejection, and validation burden.

The paper is therefore organized around four magnetic signal classes: (i) driven induction responses, (ii) leakage fields in MFL-type inspection, (iii) passive self-fields linked to stress or corrosion, and (iv) fields generated by operational currents. This structure separates two questions that are too often mixed together in the literature: which magnetic signal is physically available in a given asset, and which receiver architecture can measure that signal repeatably under realistic constraints such as stand-off, scan speed, magnetic interference, and limited access. Pipelines, reinforced concrete, steel structures, batteries, and busbars reappear across these classes as recurring use cases rather than as the primary organizing principle.

Figure~\ref{fig:intro_overview} summarizes the framing used throughout the review. The left column shows representative hidden-damage cases in which magnetic access is possible without couplants but where stand-off $h$ and clutter fields strongly influence the outcome. The top-right panel distinguishes inspection from monitoring, which clarifies why repeatability, not sensitivity alone, is the central requirement for defensible decisions. The lower panel defines the receiver-centric measurement chain adopted in the rest of the manuscript and motivates the figures of merit emphasized later: in-band noise, usable bandwidth, dynamic range, background rejection, vector/gradient capability, and explicit control of geometry, calibration, and validation. Table~\ref{tab:OPM_NV_comparison} should therefore be read as a comparison of receiver architectures inside a full measurement chain, not as a contest of isolated sensor noise floors.

\begin{table}
\centering
\small
\caption{ {Field-oriented comparison of optically pumped atomic magnetometers (OPMs) and NV-center diamond magnetometers for infrastructure NDE. The emphasis is on receiver behavior inside a measurement chain rather than on best-case stand-alone sensitivity.}}
\label{tab:OPM_NV_comparison}
\renewcommand{\arraystretch}{1.25}
{ 
\begin{tabular}{p{4.0cm} p{5.2cm} p{5.2cm}}
\hline
\textbf{Capability} & \textbf{OPM (Atomic Vapor)} & \textbf{NV-Center Diamond} \\
\hline
Primary transduction &
Optical pumping and alkali-spin precession convert magnetic field into probe-beam polarization rotation or a related optical observable &
Spin-dependent fluorescence and ODMR frequency shifts convert magnetic-field projections onto NV axes into an optical observable \\

Field-relevant operating modes &
Finite-field scalar/vector OPMs and RF-OPMs are the most practical for infrastructure; SERF is mainly a benchmark for best-case sensitivity &
Ensemble CW-ODMR point sensors, fiber-coupled heads, and compact near-surface imagers are the most relevant; pulsed protocols are useful when the signal band is known \\

Intrinsic sensitivity benchmark (not field performance) &
fT/$\sqrt{\mathrm{Hz}}$ class in shielded SERF; unshielded Earth-field or RF heads are typically in the pT/$\sqrt{\mathrm{Hz}}$ class &
Ensemble devices span pT--nT/$\sqrt{\mathrm{Hz}}$ depending on optics, microwave delivery, and readout; compact unshielded heads usually trade sensitivity for robustness \\

Best-matched infrastructure signal classes &
Driven induction responses, low-frequency susceptometry, and passive gradient/trending when strong background rejection is available &
Near-surface leakage-field mapping, vector/tensor magnetic imaging, and differential sensing of operational currents at small stand-off \\

Active region / effective sampling &
Vapor-cell volume and stand-off average the field over a mm--cm scale; spatial response is strongly shaped by cell size and scan geometry &
Set by NV layer thickness, illuminated area, and optics; sub-mm sampling is possible at small stand-off, but practical resolution is limited by spot size and sensor--target spacing \\

Usable task band &
DC--kHz is natural; RF-OPMs provide narrowband operation from kHz to MHz around a chosen bias resonance &
Near-DC to kHz is common in CW tracking; pulsed protocols target selected AC bands up to MHz \\

Dynamic-range management &
Compensation coils and closed-loop bias control determine the lock range; finite-field designs can operate in the geomagnetic field but remain sensitive to large static gradients and direct coil coupling &
Frequency-locked ODMR tracks resonance shifts directly; usable range is set by microwave tuning span, bias design, and resonance overlap at higher field \\

Vector / gradient pathway &
Multi-axis optics, arrays, and separated gradiometer heads provide vector or gradient information; common-mode rejection is often defined by hardware baseline geometry &
Four crystallographic NV orientations provide an intrinsic vector basis; tensor and gradient methods are available without mechanically rotating the sensor \\

Dominant nuisance variables &
Heading errors, light shifts, cell-temperature drift, polarization drift, direct pickup from drive coils, and magnetic-field gradients across the cell &
Temperature/strain/electric-field shifts of ODMR, laser-heating drift, microwave-field nonuniformity, resonance broadening, and added stand-off from protective packaging \\

Head-level integration burden &
Requires stable laser delivery, vapor-cell heating, magnetic bias/compensation coils, and careful thermal isolation in the scan head &
Requires stable optical pumping, efficient fluorescence collection, microwave delivery near metal, and thermal control of the diamond and its package \\

Most natural comparative role &
Strongest as a low-frequency, phase-referenced receiver when inductive coils become voltage-limited &
Strongest as a compact solid-state receiver when small stand-off, vector information, or differential current sensing is decisive \\
\hline
\end{tabular}
}
\end{table}

%%%%
\section{Principles of Quantum Magnetometry}
\label{sec:principles}

{ 
For infrastructure inspection and monitoring, a quantum magnetometer is best treated as one block inside a measurement chain. A compact expression is
\begin{equation}
y(\omega)=H_{\mathrm{read}}(\omega)\,H_{\mathrm{sens}}(\omega)\,H_{\mathrm{geom}}(\omega;h,\theta)\,B_{\mathrm{src}}(\omega)+n(\omega),
\label{eq:meas_chain}
\end{equation}
where $B_{\mathrm{src}}(\omega)$ is the source field spectrum generated by the asset or by an applied excitation, $H_{\mathrm{geom}}$ captures geometric filtering by stand-off $h$, orientation $\theta$, and spatial averaging, $H_{\mathrm{sens}}$ is the sensor transfer function, $H_{\mathrm{read}}$ is the readout or demodulation transfer function, and $n(\omega)$ is the combined technical and environmental noise seen at the output. This expression is deliberately simple, but it highlights the central practical point: in field use, the reported signal is never the source field alone.
}

{ 
Geometry is not a nuisance detail. To first order, the uncertainty associated with imperfect positioning can be written as
\begin{equation}
\delta B_{\mathrm{geom}} \approx \left|\frac{\partial B}{\partial h}\right|\delta h + \left|\frac{\partial B}{\partial \theta}\right|\delta\theta ,
\label{eq:geom_error}
\end{equation}
which is why lift-off metrology and orientation control are so often more valuable than marginal improvements in intrinsic sensitivity. The same argument applies to vector and gradient measurements: unless the sensor axes and baselines are known, the additional channels increase ambiguity rather than reduce it.
}

{ 
Several figures of merit connect sensor specifications to what is actually observed in a scan. First, the magnetic noise spectral density $n_B(f)$ (T/$\sqrt{\mathrm{Hz}}$) should be stated in the task band rather than quoted as a single best-point number, because many infrastructure signatures lie in the Hz--kHz range. Second, the effective bandwidth is the frequency range over which the \emph{combined} sensor-plus-readout system tracks variations in $\mathbf{B}$ with a known transfer function. Third, the dynamic range---or, in closed-loop systems, the lock range---must be stated in the presence of realistic background fields, including the geomagnetic field, nearby conductors, and any intentional magnetizer or bias field. Fourth, monitoring applications require a stability metric in addition to noise density. A standard choice is Allan deviation,
\begin{equation}
\sigma_B^2(\tau)=\frac{1}{2}\Big\langle\big(\bar{B}_{k+1}(\tau)-\bar{B}_{k}(\tau)\big)^2\Big\rangle ,
\label{eq:allan}
\end{equation}
because it distinguishes white-noise averaging from drift-dominated behavior and therefore shows whether longer averaging actually improves repeatability.
}

{ 
Scan dynamics set the relevant bandwidth. A spatial feature of characteristic length $\ell$ encountered at scan speed $v$ generates signal content on the order of
\begin{equation}
f_{\mathrm{sig}}\sim \frac{v}{\ell},
\label{eq:scan_band}
\end{equation}
with additional harmonics set by feature shape, sampling, and reconstruction. Faster scans and smaller defects push signal power upward in frequency, while slow drift and 50/60~Hz interference often dominate the baseline at the low-frequency end. This simple scaling already explains why narrowband, phase-referenced readout is so attractive for driven magnetic measurements.
}

{ 
Lock-in amplifiers implement phase-sensitive detection by multiplying the measured channel by in-phase and quadrature references at a known angular frequency $\omega_{\mathrm{ref}}$ and then low-pass filtering the products \cite{Scofield1994LockIn,Meade1983LockIn}. Writing the measured channel as $x(t)=A\cos(\omega_{\mathrm{ref}}t+\phi)+n(t)$, the demodulated outputs are
\begin{equation}
X=\mathrm{LPF}\!\left[2x(t)\cos(\omega_{\mathrm{ref}}t)\right],\qquad
Y=\mathrm{LPF}\!\left[2x(t)\sin(\omega_{\mathrm{ref}}t)\right],
\label{eq:lockin_xy}
\end{equation}
from which the recovered amplitude and phase follow as
\begin{equation}
A=\sqrt{X^2+Y^2},\qquad \phi=\tan^{-1}\!\left(\frac{Y}{X}\right).
\label{eq:lockin_ap}
\end{equation}
For infrastructure measurements, $X$ and $Y$ are not mere plotting choices: together they represent a complex transfer function at the drive frequency, and they can respond differently to lift-off, conductivity, permeability, and geometry.
}

{ 
For unshielded scanning, closed-loop resonance tracking is often more robust than repeated full sweeps. In tracking mode, a feedback loop continuously adjusts a control variable---for example, laser modulation frequency, microwave frequency, or a compensation field---to keep the sensor at a defined operating point on a resonance. The logged control signal then becomes the measurement output. This suppresses slow baseline wander by absorbing much of the drift inside the feedback loop rather than leaving it in the recorded time series.
}

{ 
Machine learning (ML) can strengthen this chain, but it should be viewed as part of readout or inversion rather than as a substitute for calibration. In practice, ML is most useful for three tasks: faster resonance estimation from raw optical data, nuisance-parameter estimation (for example, lift-off or orientation), and inverse mapping from magnetic observables to defect metrics when the forward model is partially known but not analytically invertible. That opportunity exists for both NV and OPM platforms \cite{Homrighausen2023EdgeMLNV,Iivanainen2022CalibrationOPM,Huang2023DeepLearningMFLReview,Zhu2019DeepLearningECT,Cormerais2022ANN_ECT}. For infrastructure use, however, any learned stage still has to be validated across the nuisance variables that dominate field performance: lift-off, alignment, background clutter, magnetization history, and scan-to-scan repeatability.
}

\subsection{Optically pumped atomic magnetometers}
\label{sec:OPMprinciples}

\begin{figure}[!htb]
\centering
\includegraphics[width=0.99\columnwidth]{  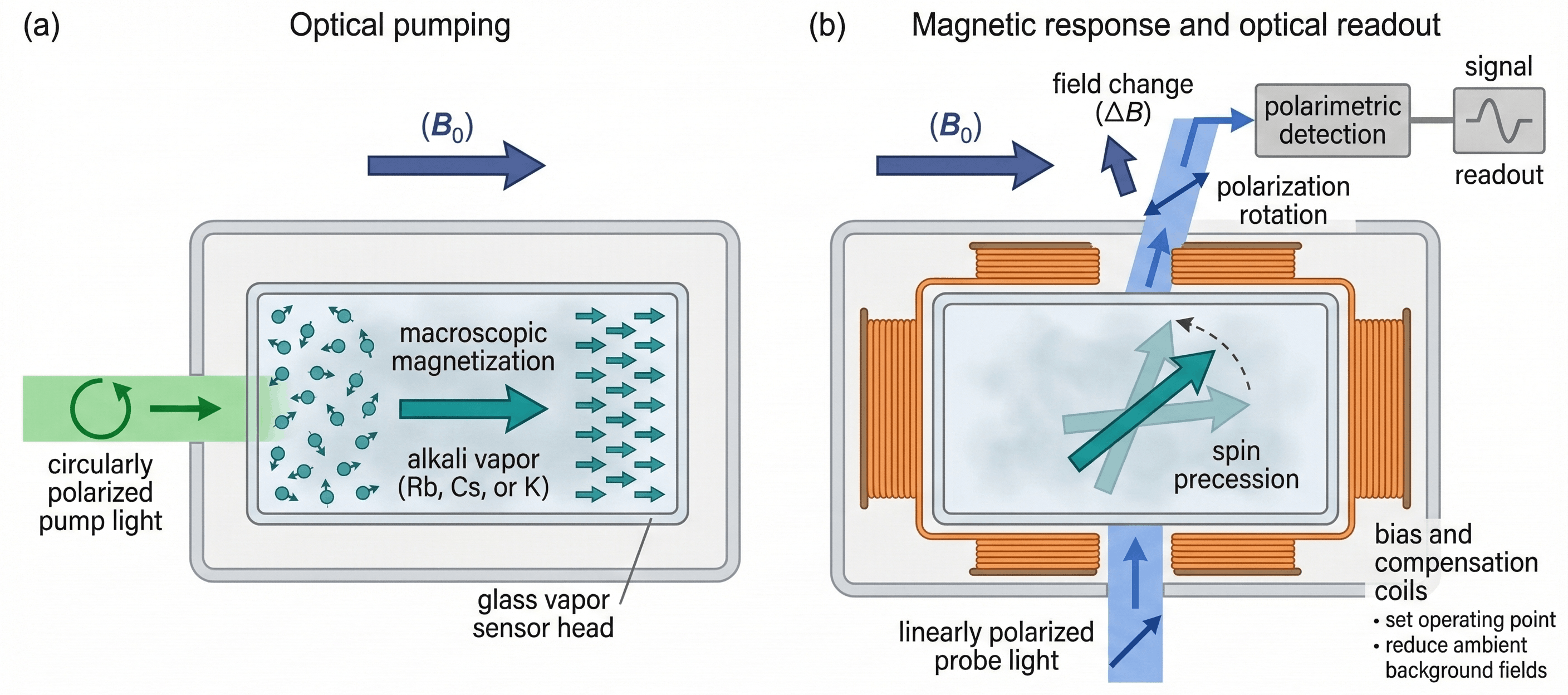}
\vspace{-9pt}
\caption{ {Operating principle of an alkali-vapor magnetometer. (a) Circularly polarized pump light spin-polarizes an alkali vapor (Rb, Cs, or K), producing a macroscopic magnetization aligned with the pump axis or a chosen bias field $B_0$. (b) Changes in magnetic field drive spin precession, which is read out optically, commonly through probe-beam polarization rotation and polarimetric detection. For field deployment, the same head typically includes bias and compensation coils that define the operating point and help reject ambient background fields.}}
\label{fig:basic_OPM}
\end{figure}

{ 
Optically pumped atomic magnetometers (OPMs) measure magnetic fields through the spin dynamics of optically polarized alkali atoms in a vapor cell \cite{BudkerRomalis2007OpticalMagnetometry,Kitching2011AtomicSensorsReview,Shah2007SubpicoteslaOPM,Dang2010UltrahighSensitivityOPM}. The sensing medium is usually a vapor of Rb, Cs, or K contained in a glass or microfabricated cell, often with buffer gas or wall coatings chosen to extend coherence time. Circularly polarized pump light transfers angular momentum to the atoms and prepares a collective spin polarization that can be described as a macroscopic magnetization $\mathbf{M}$ (A/m). A second, usually linearly polarized, probe beam reads out the spin state through optical rotation or related polarization changes. This collective-spin description is particularly useful for infrastructure sensing because it makes the OPM behave like a tunable magnetic receiver whose transfer function is set by optical pumping, relaxation, and feedback.
}

{ 
A convenient dynamical model is the Bloch form
\begin{equation}
\frac{d\mathbf{M}}{dt}
=
\gamma\,\mathbf{M}\times\mathbf{B}_{\mathrm{tot}}
-\frac{M_x\hat{\mathbf{x}}+M_y\hat{\mathbf{y}}}{T_2}
-\frac{(M_z-M_0)\hat{\mathbf{z}}}{T_1}
+\mathbf{R}_{\mathrm{pump}},
\label{eq:bloch_opm}
\end{equation}
where $\gamma$ is the atomic gyromagnetic ratio, $T_1$ and $T_2$ are the longitudinal and transverse relaxation times, $M_0$ is the steady-state polarization in the absence of perturbation, and $\mathbf{R}_{\mathrm{pump}}$ represents optical pumping. In the simplest scalar picture, the polarized spins precess at the Larmor angular frequency
\begin{equation}
\omega_L=\gamma \left|\mathbf{B}_{\mathrm{tot}}\right|,
\label{eq:larmor_opm}
\end{equation}
with
\begin{equation}
\mathbf{B}_{\mathrm{tot}}(t)=\mathbf{B}_0+\Delta \mathbf{B}(t)+\mathbf{B}_{\mathrm{bg}}(t),
\label{eq:Btotal_opm}
\end{equation}
where $\mathbf{B}_0$ is the chosen operating bias, $\Delta\mathbf{B}(t)$ is the task signal, and $\mathbf{B}_{\mathrm{bg}}(t)$ is any environmental background. The choice of $\mathbf{B}_0$ determines which field component is measured linearly and how much excursion can be tolerated before the sensor leaves its valid operating region.
}

{ 
The optical readout is usually a polarimetric one. To first order,
\begin{equation}
S(t)\approx G\,\theta_F(t),\qquad \theta_F(t)\propto M_{\perp}(t),
\label{eq:faraday_opm}
\end{equation}
where $S(t)$ is the detector output, $\theta_F$ is the probe polarization rotation, $M_{\perp}$ is the relevant transverse spin component, and $G$ is the polarimeter gain. In practice, the field dependence of $S$ is often measured through lock-in detection or through a closed-loop tracking architecture rather than by directly fitting free precession on every point of a scan.
}

{ 
The term OPM covers several operating regimes with very different field tolerance and deployment implications. Spin-exchange-relaxation-free (SERF) OPMs achieve the best-known atomic sensitivities by operating at high alkali density and near zero magnetic field, where spin-exchange broadening is strongly suppressed \cite{Allred2002SERF,Kominis2003SERF}. They remain essential reference devices, but they usually require shielding or aggressive field cancellation and therefore are not the default architecture for infrastructure work. Fieldable instruments more commonly use finite-field scalar or vector OPMs, including M$_x$/M$_z$-type or related driven modes, and radio-frequency OPMs (RF-OPMs) that are biased to respond resonantly at a chosen excitation frequency \cite{Deans2018UnshieldedRFAM,MarmugiRenzoni2020EMI,oelsner2022earthfield}. Those architectures are more relevant for scanning in the geomagnetic field or near drive coils, because the operating point can be defined and actively maintained without requiring near-zero ambient field.
}

{ 
The practical noise-equivalent field in the task band is set by both optical/electronic noise and the field-to-signal slope,
\begin{equation}
\delta B = \frac{\delta S}{\left|dS/dB\right|},
\label{eq:OPM_dB}
\end{equation}
where $\delta S$ is the signal noise inside the measurement bandwidth. The familiar best-case scaling is
\begin{equation}
\delta B_{\mathrm{min}}\sim \frac{1}{\gamma}\frac{1}{\sqrt{N\,T_2\,T}},
\label{eq:OPMsensitivity}
\end{equation}
where $N$ is the number of contributing atoms and $T$ is the averaging time \cite{BudkerRomalis2007OpticalMagnetometry,Kitching2011AtomicSensorsReview}. For deployment, however, the more relevant quantity is often the realized in-band noise under unshielded conditions, because technical noise, magnetic clutter, and direct coupling to nearby excitation coils commonly dominate before the intrinsic limit is reached.
}

{ 
Bandwidth is directly tied to relaxation. A useful decomposition is
\begin{equation}
\Gamma_2=T_2^{-1}=
\Gamma_{\mathrm{sd}}+\Gamma_{\mathrm{wall}}+\Gamma_{\mathrm{grad}}+\Gamma_{\mathrm{pump}}+\Gamma_{\mathrm{se}}^{\mathrm{eff}},
\label{eq:Gamma2_opm}
\end{equation}
where the terms denote spin-destruction, wall, field-gradient, pumping-induced, and effective spin-exchange contributions. Linearizing Eq.~\eqref{eq:bloch_opm} about the operating point gives a first-order response of the form
\begin{equation}
\Delta S(\omega)\propto \frac{\gamma M_0T_2}{1+i\omega T_2}\,\Delta B(\omega),
\label{eq:opm_transfer}
\end{equation}
so that the characteristic bandwidth scales as
\begin{equation}
f_{\mathrm{bw}}\sim \frac{1}{2\pi T_2}.
\label{eq:OPM_bandwidth}
\end{equation}
This is a central deployment trade-off: widening bandwidth by shortening $T_2$ also reduces slope and therefore degrades sensitivity. The correct operating point is therefore application dependent rather than universal.
}

{ 
Background rejection is usually implemented at the hardware level. A simple first-order gradiometer formed from two sensors separated by a baseline $d$ measures
\begin{equation}
\frac{\partial B}{\partial x}\approx \frac{B_1-B_2}{d},
\label{eq:gradiometer_opm}
\end{equation}
which suppresses common-mode geomagnetic drift and moving magnetic clutter \cite{Fu2020ChallengingEnvironments}. More elaborate arrays can recover vector or higher-order spatial information, but the benefit depends on stable channel-to-channel calibration, known baselines, and matched bandwidths.
}

{ 
For infrastructure deployment, OPM performance is shaped by several nonidealities that are easy to underestimate in laboratory demonstrations. The sensor response depends on heading relative to the bias field and optical geometry; some scalar configurations exhibit dead zones or heading errors. Light shifts, polarization drift, and cell-temperature variation change the operating point. Strong nearby coils or magnetizers can couple directly into the head or create gradients across the vapor cell. The finite cell volume also means that the OPM averages the field over a non-negligible region rather than sampling a mathematical point. Modern sensor heads reduce these problems through microfabricated vapor cells, VCSEL-based optics, compact photodiodes, integrated bias/compensation coils, and thermal isolation that confines the heated cell to the head \cite{Schwindt2004ChipScaleOPM,Kitching2018ChipScaleAtomicDevices,Alem2017OPMImaging}. Even so, the practical instrument still needs a mechanical fixture that controls stand-off $h$ and orientation, because geometry errors usually dominate the residual uncertainty.
}

{ 
In short, OPMs are best understood as tunable low-frequency magnetic receivers whose greatest comparative advantage appears when the informative signal lies in the Hz--kHz range and conventional pickup coils become voltage-limited. That is why they are especially compelling for phase-referenced induction measurements and other tasks in which the signal exists at low frequency but can still be referenced, tracked, and background-rejected.
}

\subsection{NV-center diamond magnetometers}
\label{sec:NVprinciples}

\begin{figure}[t]
\centering
\includegraphics[width=\linewidth]{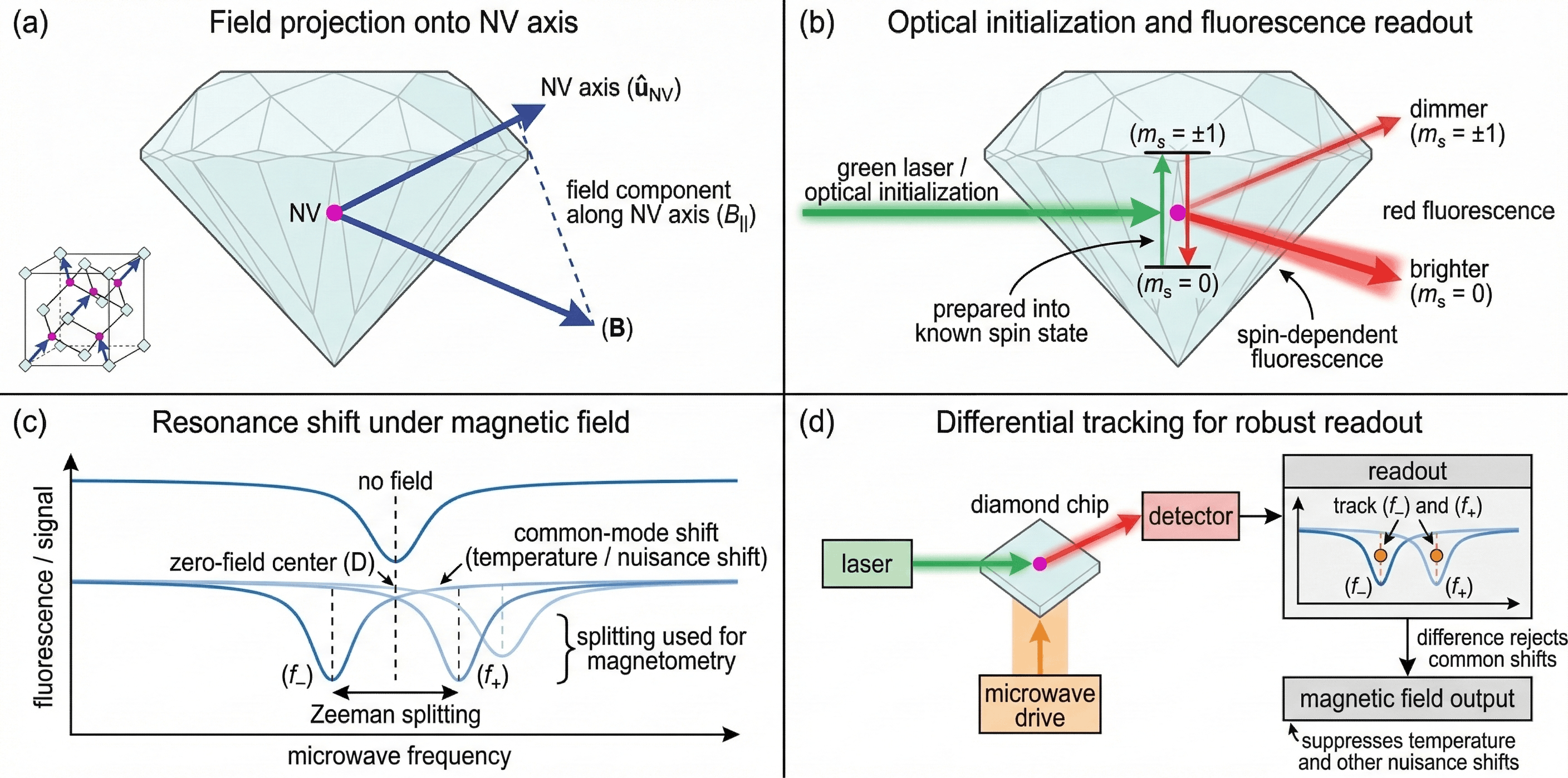}
\caption{Diamond NV-center magnetometer. (a) Magnetic-field projection onto the NV axis, so the relevant measured quantity is the axial component $B_{\parallel}=\mathbf{B}\cdot\hat{\mathbf{u}}_{\mathrm{NV}}$. (b) Optical initialization and spin-dependent fluorescence readout. Green excitation prepares the spin toward the $m_s=0$ state, while the fluorescence depends on spin state. (c) ODMR resonance shift under magnetic field: Zeeman splitting separates the resonances to $f_{-}$ and $f_{+}$, while common-mode shifts can move both resonances together. (d) Differential tracking of the split resonances provides a robust magnetic-field output while suppressing temperature-driven and other nuisance shifts.}
\label{fig:NVC_physics}
\end{figure}

{ 
Nitrogen-vacancy (NV) center magnetometers use the spin-dependent fluorescence of the negatively charged NV$^{-}$ defect in diamond to measure magnetic field at room temperature \cite{Doherty2013NVReview,Rondin2014NVReview,Degen2017QuantumSensing,Barry2020Sensitivity}. The platform spans several regimes: single-NV probes can deliver nanometer-scale sensing volumes, wide-field ensemble layers support magnetic imaging, and compact ensemble point sensors provide robust field receivers in a small solid-state package. For infrastructure applications, the most relevant architectures are ensemble point sensors, fiber-coupled heads, and compact near-surface imagers rather than single-defect nanoscale microscopy.
}

{ 
The starting point is the ground-state spin Hamiltonian
\begin{equation}
\hat{H}/h
=
D S_z^2
+
E\left(S_x^2-S_y^2\right)
+
\gamma_e\,\mathbf{B}\cdot\mathbf{S}
+
\hat{H}_{\mathrm{hf}}
+
\hat{H}_{\mathrm{strain/electric}},
\label{eq:nv_hamiltonian}
\end{equation}
where $D\approx \SI{2.87}{\giga\hertz}$ is the zero-field splitting, $E$ represents transverse splitting from strain or symmetry breaking, $\gamma_e$ is the electron gyromagnetic ratio, and the remaining terms describe hyperfine and strain/electric couplings. The simplified engineering picture follows by projecting the magnetic field onto a given NV axis $\hat{\mathbf{u}}_{\mathrm{NV}}$, so that
\begin{equation}
B_{\parallel}=\mathbf{B}\cdot\hat{\mathbf{u}}_{\mathrm{NV}}.
\label{eq:nv_projection}
\end{equation}
In bulk diamond there are four crystallographic NV orientations, which is why ensemble NV sensors can provide vector information without mechanically rotating the sensor.
}

{
Figure~\ref{fig:NVC_physics}(a) introduces the geometric basis of NV magnetometry. An NV center responds primarily to the projection of the magnetic field onto its symmetry axis, so the relevant quantity is the axial component $B_{\parallel}=\mathbf{B}\cdot\hat{\mathbf{u}}_{\mathrm{NV}}$. Figure~\ref{fig:NVC_physics}(b) then sketches the optical cycle used for readout. Green excitation, commonly at 532~nm, optically pumps the spin toward the $m_s=0$ ground state. The fluorescence depends on spin state because the excited-state relaxation includes a non-radiative intersystem crossing through a singlet manifold \cite{Doherty2013NVReview}. This spin-selective fluorescence is the basis of optically detected magnetic resonance (ODMR): microwaves drive the $m_s=0\leftrightarrow m_s=\pm1$ transitions, and the resulting fluorescence contrast is measured.
}

{
Figure~\ref{fig:NVC_physics}(c) illustrates the key spectral signature used for magnetometry. In the simple axial limit, the resonance frequencies are
\begin{equation}
f_{\pm}=D\pm \frac{\gamma_e}{2\pi}B_{\parallel},
\label{eq:nv_simple}
\end{equation}
so that the field projection can be estimated from the splitting
\begin{equation}
B_{\parallel}=\frac{\pi}{\gamma_e}\left(f_+-f_-\right).
\label{eq:NV_B_from_splitting}
\end{equation}
A useful decomposition is
\begin{equation}
\bar{f}=\frac{f_++f_-}{2},\qquad \Delta f=f_+-f_-.
\label{eq:nv_center_split}
\end{equation}
To first order, $\Delta f$ carries the magnetic information whereas $\bar{f}$ carries common-mode shifts of $D$. This matters because $D$ is temperature dependent and can also be perturbed by strain and electric fields. Near room temperature, a commonly used value is $dD/dT\approx -\SI{74}{\kilo\hertz\per\kelvin}$ \cite{Acosta2010NVTemperature}. As indicated schematically by the common-mode displacement in Fig.~\ref{fig:NVC_physics}(c), protocols that track only one ODMR line are vulnerable to thermal or strain drift, whereas differential tracking of the split resonances suppresses those common-mode contributions.
}

{
Two readout families dominate. Continuous-wave ODMR (CW ODMR) sweeps or modulates the microwave frequency while monitoring fluorescence, and it remains the most straightforward choice for mapping and field deployment \cite{Rondin2014NVReview,Barry2020Sensitivity}. In scanning systems, CW ODMR is often used in frequency-tracking mode, as illustrated schematically in Fig.~\ref{fig:NVC_physics}(d): the microwave frequency is modulated, the fluorescence is demodulated with a lock-in or related readout, and a feedback loop holds the sensor on resonance by tracking both $f_{-}$ and $f_{+}$. Pulsed protocols---for example Ramsey, Hahn-echo, or dynamical-decoupling sequences---convert magnetic field into an accumulated phase,
\begin{equation}
\phi=\gamma_e B_{\parallel}\tau ,
\label{eq:nv_phase}
\end{equation}
during a controlled free-evolution interval $\tau$. Pulsed sensing can improve sensitivity in selected AC bands and suppress some low-frequency technical noise, but it requires tighter timing, more complex microwave control, and a signal spectrum that is known well enough to exploit the sequence bandwidth.
}

The vector character of the platform is already suggested by the crystallographic inset in Fig.~\ref{fig:NVC_physics}(a). In bulk diamond there are four possible NV orientations, so vector reconstruction uses four field projections. If $B_i=\hat{\mathbf{u}}_i\cdot\mathbf{B}$ for $i=1\ldots 4$, then
\begin{equation}
\mathbf{b}=
\begin{bmatrix}
B_1\\
B_2\\
B_3\\
B_4
\end{bmatrix}
=
\mathbf{U}\mathbf{B},
\qquad
\mathbf{B}=(\mathbf{U}^{T}\mathbf{U})^{-1}\mathbf{U}^{T}\mathbf{b},
\label{eq:NV_vector_recon}
\end{equation}
where the rows of $\mathbf{U}$ are the NV unit vectors in the laboratory frame \cite{schloss2018simultaneous}. In practice, the reconstruction is only as reliable as the orientation metrology of the diamond relative to the scan frame. That orientation is fixed by crystal mounting, bias-field alignment, and any motion of the sensor head, so vector capability does not remove the need for geometric control.

For ensemble NV sensors operated in CW ODMR, a useful sensitivity estimate is
\begin{equation}
\eta_B \approx \frac{2\pi}{\gamma_e}\,\frac{\delta}{C\sqrt{R}},
\label{eq:NV_sensitivity}
\end{equation}
where $\delta$ is the ODMR linewidth, $C$ is the ODMR contrast, and $R$ is the detected photon rate \cite{Rondin2014NVReview,Barry2020Sensitivity}. In the language of Fig.~\ref{fig:NVC_physics}(b)--(d), linewidth is set by how sharp the ODMR dips remain, contrast reflects how strongly the fluorescence depends on spin state, and $R$ is determined by how efficiently the red fluorescence is collected at the detector. This is why optical collection design and compact fiber-coupled architectures matter so much for deployable heads \cite{Krasnok2015DielectricNanoantenna,Krasnok2014Superdirective,Patel2020FiberCoupledNV}.

In applied NV magnetometry, dynamic range is as important as sensitivity. Near magnetizers, busbars, or current-carrying equipment, the defect signal may be small even while the static background shifts the ODMR frequencies substantially. Closed-loop frequency-locking schemes address this by directly tracking the resonance frequencies, as illustrated in Fig.~\ref{fig:NVC_physics}(d), rather than inferring field from small fluorescence changes at a fixed microwave frequency \cite{Clevenson2018HDRVectorNV,Kumar2024MagPI}. The practical dynamic range is then set by the microwave tuning span, bias design, and the extent to which the ODMR lines remain distinguishable rather than by the narrowest open-loop linear region around a single operating point.

A portable NV head must solve a different set of engineering problems from an OPM head. The sensing element itself is solid state and room-temperature compatible, which simplifies ruggedization, but the package still requires stable optical pumping, efficient fluorescence collection, microwave delivery close to metal, and thermal control of the diamond and surrounding mount \cite{Sturner2019CompactNV,stuerner2021portable,graham2023fiber,Patel2020FiberCoupledNV}. The compact readout chain in Fig.~\ref{fig:NVC_physics}(d) emphasizes this system view: laser excitation, microwave drive, fluorescence detection, and resonance tracking are all part of the measurement head. Additional stand-off may be introduced by a protective cap, adhesive layer, or window between the diamond and the asset, and that spacing must be counted as part of the geometry model. The main nuisance variables are therefore temperature, strain/electric-field shifts, microwave nonuniformity, resonance broadening, and imperfect knowledge of the sensor-to-target geometry.

In comparative terms, NV magnetometers are strongest when the task rewards small stand-off, vector or gradient information, or a compact solid-state sensing head. That makes them particularly attractive for near-surface leakage-field mapping, tensor or gradient measurements, and differential sensing of operational currents, even though the underlying defect or current reconstruction remains constrained by geometry and inverse-problem assumptions.

%%%%%%%%%%%%%%%%%%
\section{ {Applications to Infrastructure Health Monitoring}}
\label{sec:apps}

{ 

Table~\ref{tab:OPM_NV_comparison} compares receiver architectures, but infrastructure performance is set by the full measurement chain rather than by the sensing element alone. In field use, an optically pumped magnetometer (OPM) or a nitrogen-vacancy (NV) diamond magnetometer reports a magnetic flux density $B(t)$ at a particular position and orientation. The asset physics determines which magnetic signal actually exists; stand-off, scan kinematics, magnetic clutter, and readout strategy determine how much of that signal survives into a repeatable map. For that reason, this section follows the \emph{signal class} rather than the asset class wherever possible, because the source physics sets the relevant bandwidth, dynamic range, background-rejection strategy, and calibration burden \cite{Wang2012ReviewThreeMagneticNDT,GarciaMartin2011ECTReview,Ma2021PipelineILI}.

The present part treats the first three signal classes introduced in Sec.~\ref{sec:intro}: driven induction responses, leakage fields in magnetic flux leakage (MFL) inspection, and passive self-fields linked to stress or corrosion. Operational-current fields are treated later in Sec.~\ref{sec:batteries}. This ordering matters because each class fails for different reasons in the field. Driven induction measurements fail primarily through poor primary/secondary-field separation, lift-off uncertainty, and phase instability. Leakage-field measurements fail through uncontrolled magnetization state and stand-off. Passive self-field methods fail through magnetic history, environmental drift, and ambiguity of interpretation. Across all three classes, deployability depends on controlling the variables to which the signal is most sensitive and on reporting calibration and validation in a form that supports an engineering decision rather than an isolated image.
}

\subsection{Driven induction responses: hidden corrosion and metal loss}
\label{sec:corrosion}

{ 
Driven induction responses are the clearest setting in which to compare quantum receivers with conventional inductive pickup, because the source physics is externally imposed and the informative signal is referenced to a known excitation \cite{GarciaMartin2011ECTReview,Deans2016ElectromagneticInductionRFAM,Wickenbrock2016EddyCurrentRFAM}. The price of that causal clarity is that the defect-sensitive secondary field is usually much smaller than the directly applied primary field, so primary-field rejection, dynamic range, and phase stability are as important as raw sensitivity.
}

\paragraph{Electromagnetic induction imaging under insulation.}

\begin{figure}[!htb]
\centering
\includegraphics[width=0.95\columnwidth]{  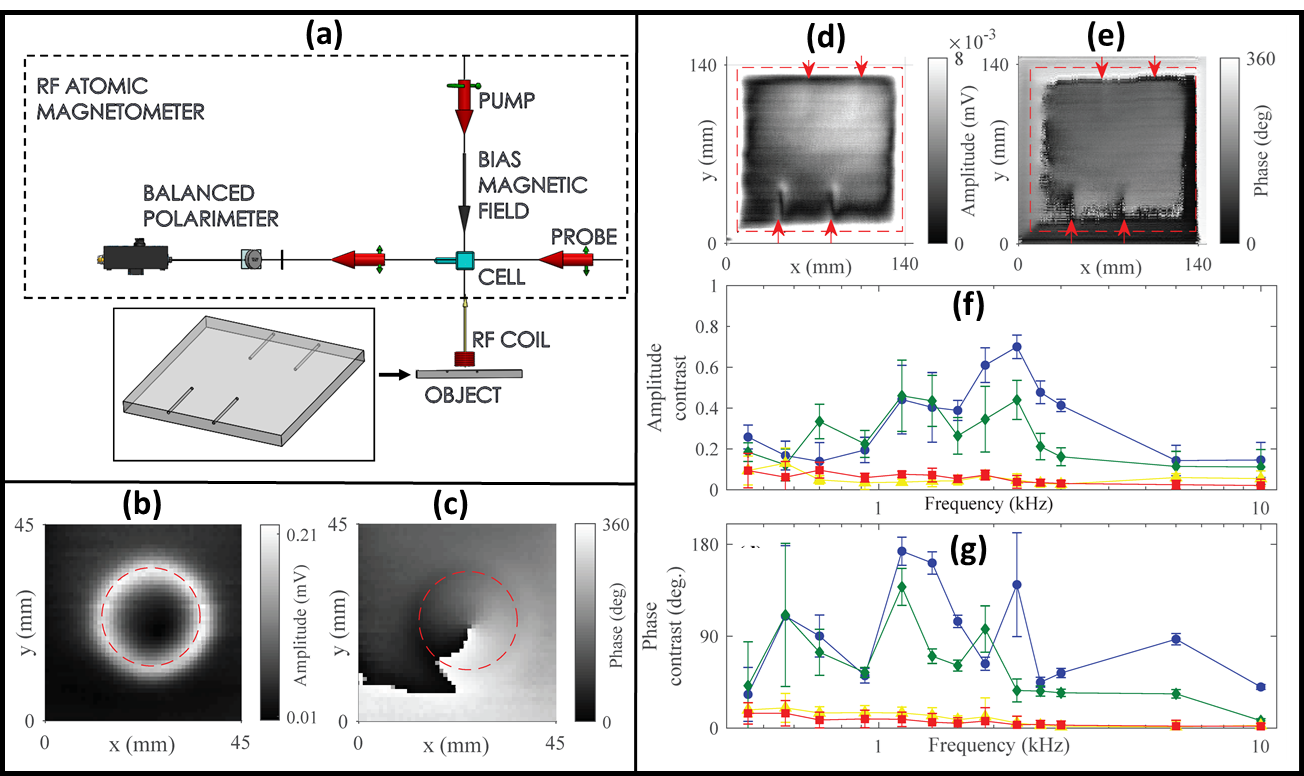}
\caption{Atomic-magnetometer electromagnetic induction imaging (EMI). (a) A drive coil generates the primary field and induced eddy currents; the magnetometer detects the secondary magnetic response. (b--e) Representative lock-in amplitude and phase maps over aluminum plates with concealed recesses and cavities. (f,g) Frequency-dependent contrast for different defect depths, illustrating tunable depth sensitivity through the choice of excitation frequency \cite{bevington2020inductive}.}
\label{fig:OPM_CUI_EMI_results}
\vspace{-6pt}
\end{figure}

{ 
Corrosion under insulation (CUI) is fundamentally a screening problem. Insulation, cladding, and coatings hide early wall loss and add uncertain stand-off, so the practical workflow is usually layered: risk-based targeting, rapid areal screening, and then higher-specificity follow-up or selective insulation removal on flagged zones. Established baselines include pulsed eddy current testing (PECT), guided-wave ultrasonics, radiography in limited circumstances, and direct insulation removal \cite{API_RP_583,DNV_RPG109,Cao2022CUIReview,Sophian2017PECTReview}. A new receiver matters only if it improves the decision process under the lift-off variability, access limitations, and magnetic interference typical of real sites.
}

Atomic-magnetometer EMI is a driven measurement in which the magnetometer reads the magnetic field of induced currents rather than the voltage induced in a pickup coil. The measured narrowband field can be written schematically as

\begin{equation}
B_{\mathrm{meas}}(\omega)=B_{\mathrm{p}}(\omega)+B_{\mathrm{s}}(\omega)+B_{\mathrm{bg}}(\omega),
\label{eq:emi_meas}
\end{equation}
where $B_{\mathrm{p}}$ is the directly applied primary field, $B_{\mathrm{s}}$ is the secondary field from induced currents in the asset, and $B_{\mathrm{bg}}$ is environmental background. The quantity of interest is often the complex response

\begin{equation}
\chi(\omega)=\frac{B_{\mathrm{s}}(\omega)}{B_{\mathrm{p}}(\omega)}=|\chi(\omega)|e^{i\phi(\omega)},
\label{eq:emi_complex}
\end{equation}
because the amplitude $|\chi|$ and phase $\phi$ respond differently to lift-off, conductivity, magnetic permeability, and geometry. This is exactly why lock-in amplitude and phase should be treated as a complex transfer observable rather than as two unrelated image channels.

Figure~\ref{fig:OPM_CUI_EMI_results}(a) shows the essential geometry: the drive coil supplies a known excitation, and the magnetometer senses the secondary field directly, without the explicit $\omega$ penalty associated with inductive voltage pickup \cite{GarciaMartin2011ECTReview}. Figures~\ref{fig:OPM_CUI_EMI_results}(b--e) show representative lock-in amplitude and phase maps over aluminum specimens containing concealed recesses and cavities \cite{bevington2020inductive}. Figures~\ref{fig:OPM_CUI_EMI_results}(f,g) show the key practical control parameter: changing the excitation frequency changes the induced-current distribution with depth and therefore changes the defect contrast.

\begin{figure}[!t]
\centering
\includegraphics[width=0.99\columnwidth]{  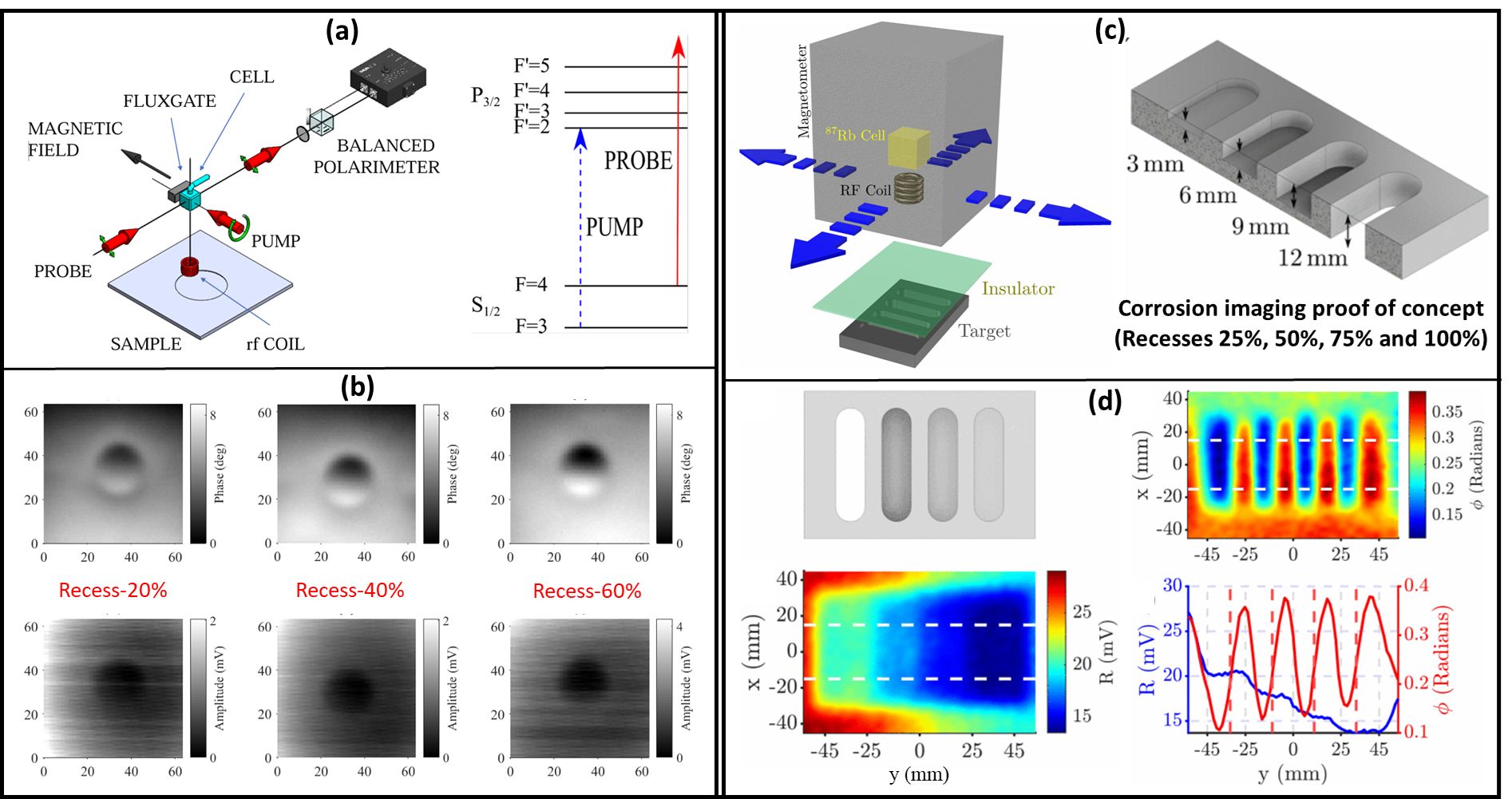}
\vspace{-9pt}
\caption{Atomic-magnetometer EMI of concealed metal loss. (a,b) Steel plate measurements showing phase and amplitude contrast for recesses of increasing depth \cite{Bevington2018Steelwork}. (c,d) Insulated aluminum specimen with recesses and a through-cut defect, with frequency-dependent contrast used to tune depth sensitivity \cite{maddox2022imaging}.}
\label{fig:OPM_CUI_EMI_recess_results}
\vspace{-6pt}
\end{figure}

A standard depth scale for induction sensing is the skin depth

\begin{equation}
\delta=\sqrt{\frac{2}{\mu\sigma\omega}},
\end{equation}
 
where $\sigma$ is the electrical conductivity and $\mu$ is the magnetic permeability. This expression is most useful as a design guide, not as a complete inversion model. In ferromagnetic steels, $\mu$ is state dependent: it changes with magnetization history, stress, microstructure, and the local field created by the drive. In CUI geometries, insulation and cladding further add lift-off uncertainty and geometric filtering. For that reason, field-oriented workflows usually rely on multi-frequency data, calibration libraries, or forward models matched to the real insulation stack rather than on a single-frequency closed-form inversion \cite{GarciaMartin2011ECTReview,Deans2016ElectromagneticInductionRFAM}.

{ 
The separation into amplitude and phase is not just a plotting convenience. The phase channel captures the time lag between excitation and response and can remain more repeatable than raw amplitude when the reference is well controlled. The amplitude channel is often the most intuitive for screening, but in ferromagnetic materials it can mix conductivity-driven and permeability-driven changes. This is precisely the regime in which OPM receivers are attractive: they preserve sensitivity at low frequency, where coil voltage pickup becomes progressively smaller, while still supporting phase-referenced measurement at a chosen drive frequency \cite{Wickenbrock2016EddyCurrentRFAM,Deans2016ElectromagneticInductionRFAM}.
}

{ 
Figure~\ref{fig:OPM_CUI_EMI_recess_results}(a,b) shows phase and amplitude responses to recesses of increasing depth in steel \cite{Bevington2018Steelwork}. Figure~\ref{fig:OPM_CUI_EMI_recess_results}(c,d) adds insulation over an aluminum specimen containing recesses and a through-cut defect; the contrast is again frequency dependent \cite{maddox2022imaging}. In both demonstrations, receiver sensitivity is only one part of performance. Coil placement, direct primary-field coupling, sensor height, scan speed, and the demodulation time constant all shape the realized transfer function. In field deployment those variables often dominate repeatability.
}

{ 
Three translation gaps recur in CUI-style deployment. First, stand-off must be treated as a measured input rather than a nuisance parameter if the goal is quantification rather than flagging. Second, inversion must be tied to realistic layered geometry and to material variability, especially in ferromagnetic steel where permeability is nonlinear. Third, performance should be reported using metrics that connect to qualification practice, such as probability-of-detection (POD) studies on representative mockups with controlled defect sets, insulation thickness, and lift-off variability \cite{ASTM_E2862_23,MIL_HDBK_1823_ASSIST}. Stated plainly, the deployable advantage of a quantum receiver in this class is not that it changes induction physics, but that it can make low-frequency, phase-referenced measurements more usable when conventional receivers become bandwidth- or noise-limited.
}

\subsection{Leakage fields in ferromagnetic structures: MFL and defect imaging}
\label{sec:cracks}

\begin{figure}[!t]
\centering
\includegraphics[width=0.87\columnwidth]{  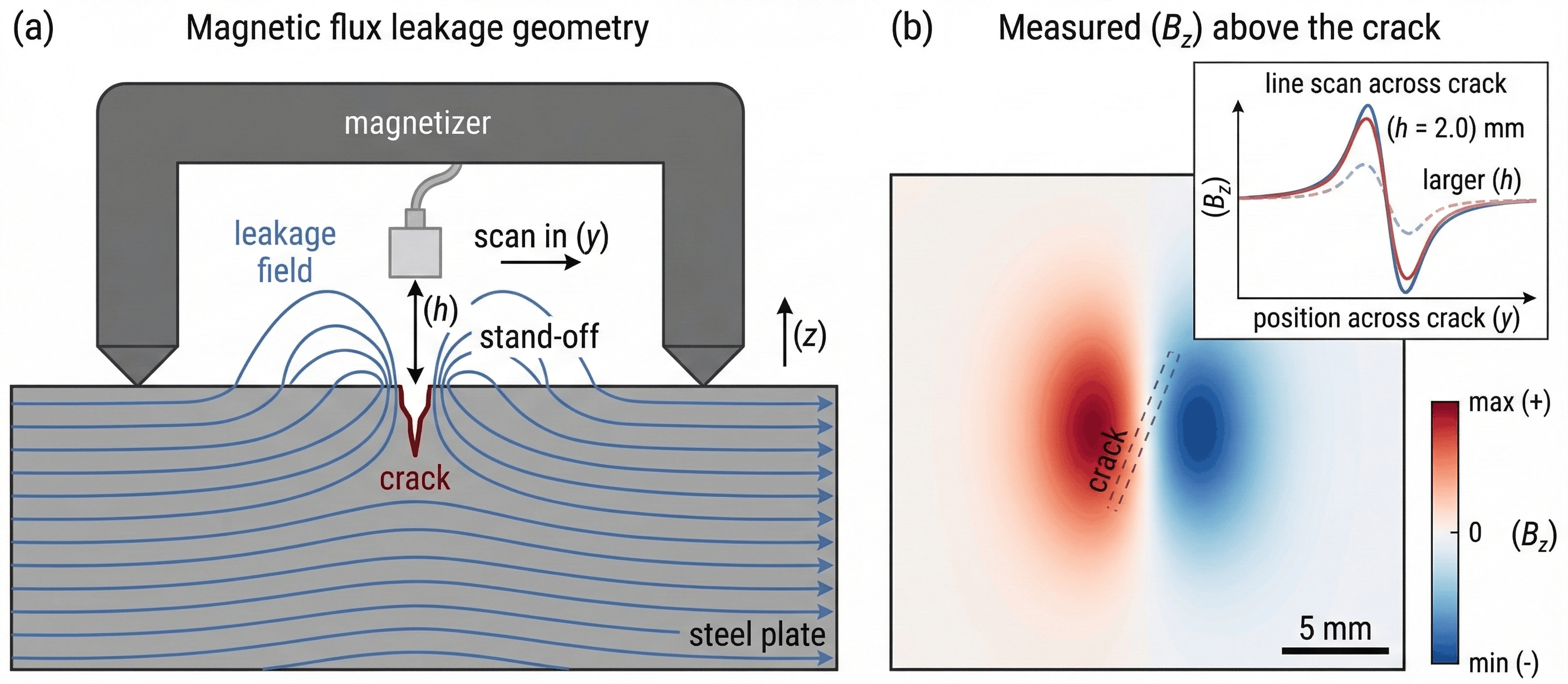}
\vspace{-9pt}
\caption{Representative stray-field geometry for a crack in magnetized steel as used in magnetic flux leakage (MFL) inspection. (a) A magnetizer drives flux through the steel; a defect perturbs the flux path and produces a leakage field above the surface, measured at a stand-off (lift-off) distance $h$ while scanning in $y$. (b) Example out-of-plane magnetic flux density component $B_z$ at $h=2.0$~mm (scale bar: 5~mm) with an inset line scan across the crack. The measured amplitude and apparent width are strongly lift-off dependent.}
\label{fig:MFL_NVC_metal_crack_sense}
\vspace{-8pt}
\end{figure}

{ 
Leakage-field inspection should be kept separate from driven induction because the source field is magnetostatic or quasi-static rather than inductively generated. In MFL, a magnetizer first establishes a flux distribution in ferromagnetic steel. A defect then appears only through the perturbation of that pre-existing flux path. The receiver therefore inherits all uncertainty associated with magnetizer geometry, local permeability, hysteresis, defect orientation, and stand-off \cite{Sun2013MagneticMechanismsMFL,Feng2022MFLReview,Ma2021PipelineILI}. This is why a more sensitive magnetic receiver does not automatically produce a better MFL instrument unless magnetization and geometry are controlled just as carefully.
}

{ 
The usual qualitative statement that ``the defect causes flux to leak into air'' is correct but incomplete. What matters for measurement is the spatial spectrum of the leakage field above the surface. For a dominant spatial wavenumber $k$, the field amplitude at stand-off $h$ is well approximated by
}
\begin{equation}
B_{\mathrm{leak}}(h;k)\approx B_{\mathrm{leak}}(0;k)e^{-kh},
\label{eq:mfl_liftoff}
\end{equation}
{ 
which makes the lift-off problem explicit: larger stand-off reduces amplitude and broadens the apparent feature, directly affecting both detectability and sizing \cite{Feng2022MFLReview}. Figure~\ref{fig:MFL_NVC_metal_crack_sense}(a) highlights the variables that matter in practice---magnetizer geometry, scan direction, sensor orientation, and stand-off distance $h$---while Fig.~\ref{fig:MFL_NVC_metal_crack_sense}(b) shows a representative $B_z$ map and line scan across a crack.
}

{ 
A second point is equally important for deployment: the magnetization state is part of the inspection method. Leakage-field signatures depend on whether the steel is near saturation, on how the magnetizer yoke and poles are arranged, and on the magnetic history of the asset \cite{Sun2013MagneticMechanismsMFL,Ma2021PipelineILI}. This is one reason MFL procedures are often more repeatable in purpose-built in-line inspection tools than in ad hoc scanning geometries. The receiver does not remove that source-physics burden.
}

\begin{figure}[t]
\centering
\includegraphics[width=\linewidth]{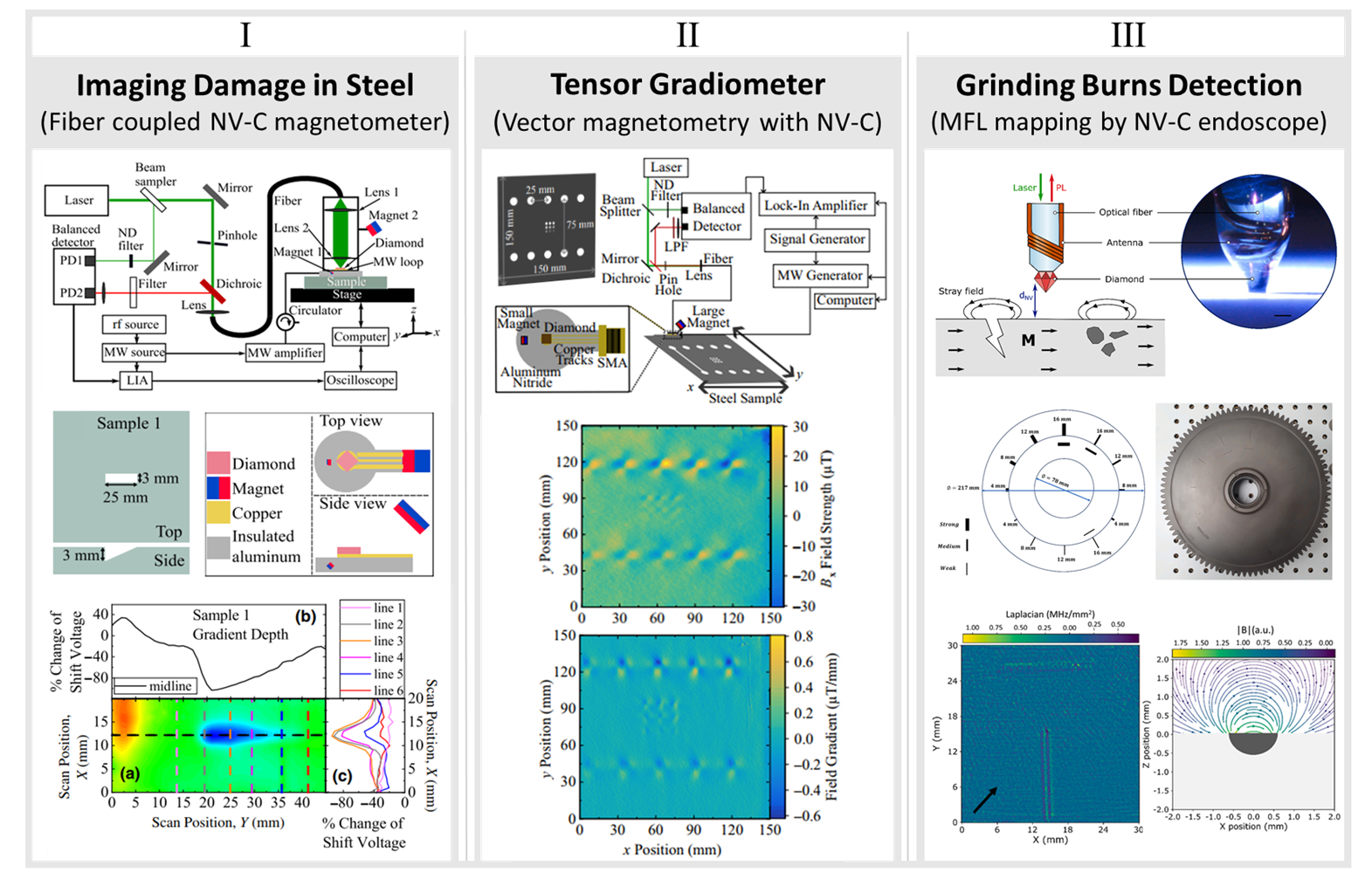}
\caption{Experimental NV-center (NV-C) imaging of metal damage and defects. (I) Fiber-coupled NV-C magnetometer used to image damage in steel by monitoring shifts of the NV Zeeman splitting caused by distortion of an intentionally inhomogeneous magnetic field; the approach operated without magnetic shielding, worked through nonmagnetic cover layers, and demonstrated mm-scale lateral defect imaging at lift-off distances up to a few millimeters \cite{zhou2021imaging}. (II) Scanning fiber-coupled NV-C tensor gradiometer with feedback-based tracking of all four NV orientations, enabling vector magnetic-field imaging and real-time tensor gradiometry for improved detection of small damage features \cite{newman2024tensor}. (III) NV-C endoscope for magnetic-flux-leakage-style detection of grinding burns in steel, combining compact optical-fiber delivery, near-surface mapping, and derivative-based post-processing to isolate burn-related magnetic signatures in industrially relevant components \cite{vindolet5186260high}.}
\label{fig:NVC_metal_crack_sensing_results}
\end{figure}

In the current room-temperature quantum literature, the most developed leakage-field implementations are NV-based rather than OPM-based, largely because compact solid-state heads can be brought close to the steel while preserving vector or gradient information. Figure~\ref{fig:NVC_metal_crack_sensing_results}(I) shows the fiber-coupled architecture reported by Zhou \emph{et al.}, where optical delivery was mechanically decoupled from the sensing head and the measurement relied on damage-induced distortion of an intentionally inhomogeneous magnetic field generated by a small permanent magnet \cite{zhou2021imaging}. Rather than seeking a highly uniform bias field on the diamond, that work used the spatial variation of the applied field itself as part of the imaging contrast mechanism. Damage in 316 stainless steel was reconstructed by quantifying shifts in the NV Zeeman splitting, with reported spatial resolution of about 1~mm parallel to the surface and 0.1~mm normal to the surface. Importantly for infrastructure-style problems, the method was demonstrated without magnetic shielding, remained functional at lift-off distances up to 3~mm, and could image damage even through nonmagnetic cover layers \cite{zhou2021imaging}.

Figure~\ref{fig:NVC_metal_crack_sensing_results}(II) shows the next step taken by Newman \emph{et al.}: a scanning fiber-coupled NV vector magnetometer with feedback control of the microwave excitation frequency to maintain sensitivity while the head moves \cite{newman2024tensor}. In that work, the frequency shifts of all four NV orientations were tracked so that the full vector magnetic field of a damaged steel plate could be reconstructed rather than a single projection alone. The resulting data stream was then used to compute tensor gradiometry images in real time. This is more than a cosmetic upgrade from scalar imaging. By combining vector information with spatial derivatives, the method suppresses slowly varying backgrounds and highlights spatially localized perturbations, allowing smaller damage to be detected than would be possible from scalar or even vector magnetic maps alone \cite{newman2024tensor}. For field deployment, the significance is that the sensor head, the tracking loop, and the spatial derivative calculation all become part of the measurement chain.

Figure~\ref{fig:NVC_metal_crack_sensing_results}(III) shows a different but highly relevant use case: the NV endoscope developed by Vindolet \emph{et al.} for high-resolution, non-destructive detection of grinding burns in steel \cite{vindolet5186260high}. Here the target was not a macroscopic crack but a thermally altered near-surface region whose magnetic signature is superimposed on the broader magnetization pattern of the part. The compact optical-fiber head enabled close access to the surface, while lock-in-assisted resonance tracking was used to map the magnetic field projected along the selected NV axis. The study demonstrated detection of all burn severities on a 16NiCrMo13 gearwheel, reported that a preliminary magnetization of about 20~mT was already sufficient to detect weak burns, and acquired 10$\times$10~mm measurement areas in minutes with effective measurement times on the order of 1~s \cite{vindolet5186260high}. The authors also showed why lift-off is a first-order variable rather than a nuisance detail: signal amplitude decreased and apparent width broadened with increasing stand-off, with their setup allowing clear quantitative detection below roughly 300~\textmu m and losing unambiguous detectability above roughly 500~\textmu m \cite{vindolet5186260high}. To further isolate the burns from the smooth background gradient between the magnetizing poles, they used derivative-based post-processing, including a Laplacian map, to emphasize sharp local changes in the magnetic field. This is a good example of a practical trade-off seen throughout NV leakage-field work: compact hardware and near-surface access improve localization, but interpretation still depends on magnetization procedure, lift-off control, and processing choices.

These examples are promising precisely because they illustrate the measurement-chain issues rather than hiding them. First, dynamic range near magnetizers is critical: even if the defect signal is small, the local background field may be large and spatially varying, so the resonance-tracking hardware must remain locked over the full operating field excursion. Second, vector and gradient channels do not eliminate calibration requirements; they intensify them. A gradient estimate depends on baseline geometry, matched channel response, and stable knowledge of the sensor pose. Third, derivative-based processing sharpens features but also amplifies noise and geometry errors.

For leakage-field inspection, OPMs are not irrelevant, but the class is less forgiving for them than driven induction because strong static backgrounds and field gradients near magnetizers can push the atomic resonance away from its preferred operating region unless compensation is carefully engineered. In contrast, NV heads are naturally compatible with close stand-off, compact packaging, and vector readout, which is why they currently appear more often in room-temperature MFL-style demonstrations. In either platform, however, quantification still requires an encoded scan head, measured stand-off, and calibration on reference artifacts representative of the steel thickness, magnetizer configuration, and defect family of interest \cite{Feng2022MFLReview,Ma2021PipelineILI}.

\subsection{Passive self-fields linked to stress or corrosion}
\label{sec:MMM}

{ 
Among the four signal classes considered in this review, passive self-fields presently have the weakest causal interpretability and the strongest sensitivity to magnetic history, lift-off, and environmental clutter. That does not make them useless; it makes them procedurally demanding. The literature in this class actually contains two different mechanisms that should not be collapsed into one category: current-generated fields from ongoing electrochemical activity, and residual or self-magnetic leakage fields produced by remanence, permeability redistribution, stress concentration, or corrosion damage in ferromagnetic materials \cite{Wang2012ReviewThreeMagneticNDT,Su2024MMMCivilEngineering}. The measurement and validation strategy is different for each.
}

\paragraph{ {Corrosion-current and remanence mapping.}}

{ 
If the measured field truly arises from an active corrosion current distribution $\mathbf{J}(\mathbf{r})$, then the forward problem is governed by the Biot--Savart relation
}
\begin{equation}
\mathbf{B}(\mathbf{r})=
\frac{\mu_0}{4\pi}
\int
\frac{\mathbf{J}(\mathbf{r}')\times\left(\mathbf{r}-\mathbf{r}'\right)}
{\left|\mathbf{r}-\mathbf{r}'\right|^3}
\,d^3r'.
\label{eq:corrosion_biot}
\end{equation}
{ 
That relation is valuable because it forces the interpretation problem into the open: the measured map is meaningful only if the current path, stand-off, and sensor axis are known well enough to support a forward model. Controlled electrochemical studies have shown that magnetic fields from corrosion-related currents can indeed be measured, first with SQUID receivers and more recently with sensitive room-temperature magnetometers \cite{Ma2002CorrosionSQUID,Bautin2021PittingCorrosionMagnetometer}. The challenge is not whether such fields exist, but whether they remain distinguishable from other quasi-static magnetic signatures in realistic steel or reinforced-concrete assets.
}

{ 
That distinction is central. In steel and reinforced concrete, a static stray-field map can arise from remanent magnetization, stress-induced changes in permeability, corrosion products, or true corrosion-current loops. Those mechanisms can produce qualitatively similar single-component maps if geometry is poorly controlled \cite{Zhang2016SteelCorrosionRCMicroMagnetic,Yang2019SMFLSteelBars,Xia2018SMFLSteelStrands}. A practical measurement plan therefore needs two things from the start: background rejection, typically through a gradiometric or reference-channel layout, and geometry metadata, namely recorded stand-off and sensor orientation. Without those metadata, it is difficult to compare scans across days or to fit a forward model that yields a physically meaningful current estimate.
}

{ 
The most defensible near-term workflow for magnetic corrosion-current mapping is therefore targeted follow-up rather than blind screening. A conventional method localizes a suspect area. A magnetic map is then acquired with controlled and recorded sensor height and orientation, ideally together with an imposed or independently measured electrochemical condition. The resulting field is fit to a small family of current-loop or distributed-current models. This forces an explicit interpretation and helps separate current-generated fields from remanence or permeability-related anomalies that would otherwise be easy to over-interpret.
}

{ 
NV ensemble heads are attractive for such follow-up measurements when small stand-off can be maintained and vector information is helpful in separating localized anomalies from broad background gradients. The main caveat is unchanged: ODMR shifts also respond to temperature, strain, and electric fields, so passive corrosion studies must use a readout protocol that isolates the magnetic contribution, such as symmetric splitting or differential tracking. Compact and fiber-coupled NV heads reduce the burden of bringing optics and microwaves into restricted geometries, but they do not remove the need for controlled stand-off and model-based interpretation \cite{stuerner2021portable}.
}

\paragraph{ {Metal magnetic memory and related self-field methods.}}

{ 
Metal magnetic memory (MMM), and related self-magnetic-flux-leakage methods, scan residual fields under ambient geomagnetic bias and infer stress concentration or damage indicators from spatial features such as extrema, zero crossings, and gradients. Their appeal is obvious: no external magnetizer is required, field deployment is simple in principle, and the measured field is quasi-static. The difficulty is equally obvious: once the excitation is no longer controlled, the result becomes much more sensitive to magnetic history, geomagnetic orientation, nearby ferromagnetic clutter, temperature drift, and stand-off \cite{Wang2012ReviewThreeMagneticNDT,Su2024MMMCivilEngineering,ISO24497_1,ISO24497_2}. That is the central trade-off of the class.
}

\begin{figure}[!t]
\centering
\includegraphics[width=0.99\columnwidth]{  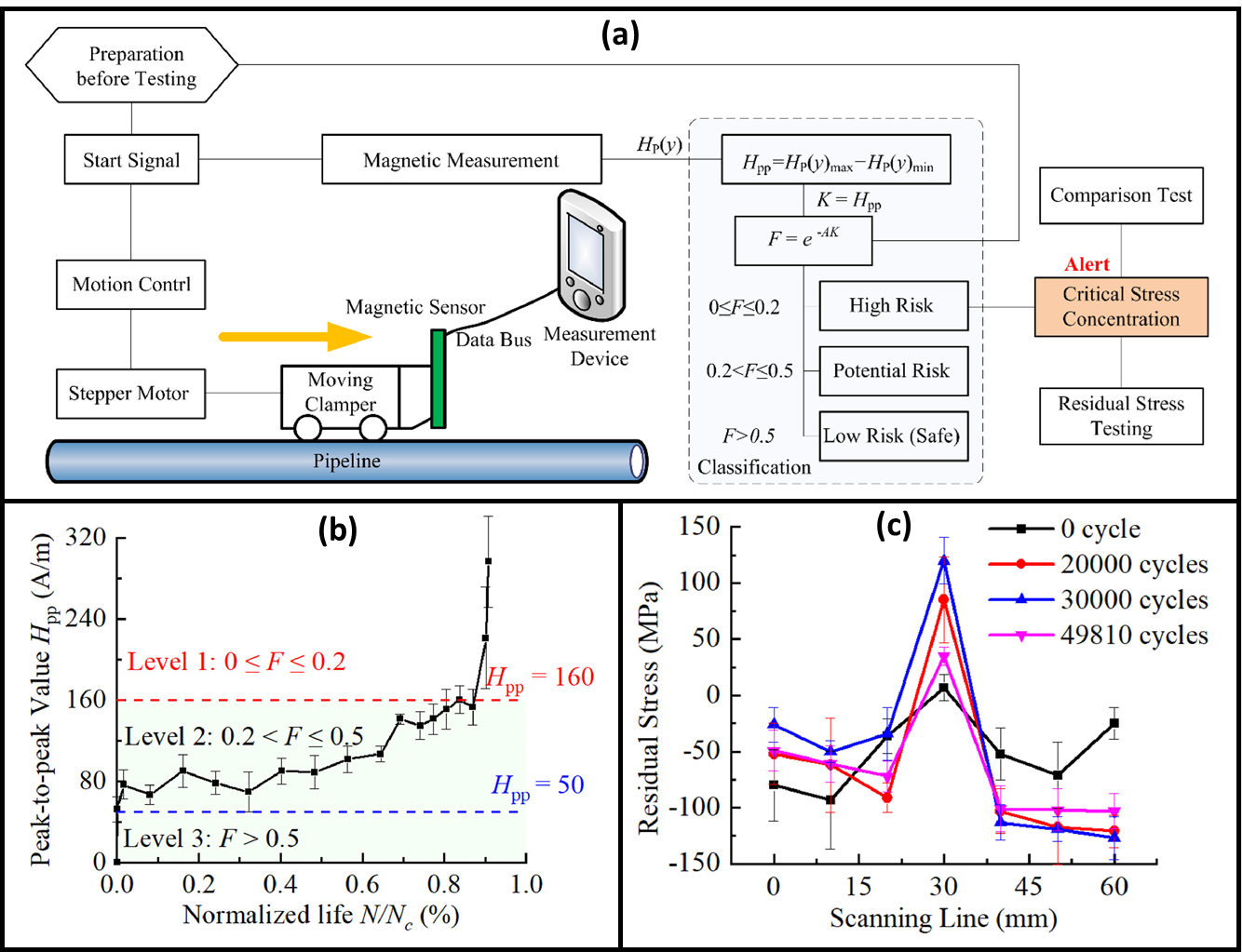}
\caption{Metal magnetic memory (MMM)--based stress-concentration diagnostics in ferromagnetic pipelines. (a) Schematic of a MMM inspection system and automated stress-concentration evaluation workflow. (b) Evolution of MMM signals and $H_{\mathrm{pp}}$ with increasing fatigue cycles, showing the progressive amplification of magnetic anomalies at stress-concentration zones. (c) Residual stress distributions along the specimen surface at different fatigue stages, illustrating the correlation between MMM signal evolution and local residual-stress changes near the defect region \cite{gong2025design}.}
\label{fig:MMM_results}
\end{figure}

Figure~\ref{fig:MMM_results}(a) shows a complete MMM workflow packaged as a practical pipeline inspection system rather than as a stand-alone laboratory scan \cite{gong2025design}. In that work, Gong \emph{et al.} built an intelligent stress-concentration detection platform around high-sensitivity anisotropic magnetoresistive (AMR) sensors (HMC5883L), an embedded S3C2140 ARM processor, an embedded Linux software stack, and evaluation algorithms implemented in a form consistent with GB/T 35090-2018. The reported hardware sensitivity was 0.4~A/m, and the system was designed to identify stress-concentration zones automatically from surface magnetic anomalies rather than requiring manual interpretation of raw magnetic traces. The scan produces a measured profile $H_P(y)$, where $H_P$ is the measured magnetic field-strength component and $y$ is the scan coordinate. A common scalar reduction is the peak-to-peak indicator
\begin{equation}
H_{\mathrm{pp}} = H_P(y)_{\max}-H_P(y)_{\min},
\end{equation}
which is then mapped to a stress-severity index or classification output. Making this reduction explicit is important for translation, because MMM is otherwise too often presented only as a field profile without a clear route to a decision metric.

The significance of Gong \emph{et al.} is not only that they measured magnetic anomalies, but that they tied those anomalies to an instrumented decision workflow. Their system introduced a magnetic anomaly evaluation index $F$ and reported a threshold $K=160$ that reliably separated plastic-deformation zones from surrounding material in their validation protocol \cite{gong2025design}. In fatigue tests, the system provided an early warning about 3200 cycles before final failure, corresponding to roughly 98.3\% of fatigue life consumed. This is exactly the kind of result that moves MMM from a qualitative inspection concept toward a structured monitoring workflow, even though the calibration remains specimen and protocol dependent.

Figure~\ref{fig:MMM_results}(b) shows how the MMM signal and the scalar feature $H_{\mathrm{pp}}$ evolve through fatigue life, and Fig.~\ref{fig:MMM_results}(c) compares this behavior with independently measured residual stress \cite{gong2025design}. The residual-stress data are particularly useful because they show that the magnetic anomaly does not merely drift arbitrarily with cycling. Before loading, the specimen surface exhibited compressive residual stress, with typical values fluctuating between about $-100$ and 0~MPa due to machining history. Away from the stress-concentration zone, the residual stress remained broadly compressive during cycling, approximately in the range $-120$ to $-20$~MPa. In contrast, the notch-affected central region underwent a compressive-to-tensile transition, with tensile residual stress growing progressively and reaching roughly +150~MPa at 48{,}500 cycles \cite{gong2025design}. That comparison makes the paper more convincing than a purely magnetic report: the magnetic anomaly is linked to an independently measured mechanical consequence of localized plasticity.

These experimental results also sit naturally within the broader MMM modeling literature. Wang \emph{et al.} addressed one of the long-standing weaknesses of MMM interpretation by using a linear magnetic-charge model to analyze the self-magnetic-flux-leakage (SMFL) distribution in a local stress-concentration zone \cite{Wang2010QuantitativeMMMStress}. Their analysis reproduced basic experimental features of MMM signals and, importantly, gave quantitative predictions for how defect depth and defect location---in particular, whether the defect was surface breaking or embedded---change the observed SMFL profile. That work did not make MMM universally quantitative, but it did show that the measured waveform is not arbitrary and that some aspects of defect geometry can, in principle, be encoded in the signal. Shi \emph{et al.} pushed the modeling further by introducing a nonlinear magnetomechanical constitutive relation under a weak ambient magnetic field and solving the resulting problem with finite-element analysis \cite{Shi2017MagnetomechanicalMMM}. Their model was benchmarked against experimental data and used to study the effects of load magnitude, defect size, specimen size, and lift-off on the magnetic-memory signal. This matters directly for field interpretation. It implies that a measured MMM anomaly depends not only on stress concentration itself, but also on the stand-off at which the sensor is held and on the overall geometry of the component. In other words, even when a useful correlation exists, the signal is filtered by the same geometry and measurement-chain factors emphasized throughout this review.

Viewed positively, MMM offers low procedural burden because no deliberate magnetizer is required. Viewed critically, that same convenience removes control of the excitation and therefore weakens causal interpretation. The most defensible use case is controlled trending on the same asset with repeatable scan geometry and a documented baseline. The least defensible use case is a universal one-time diagnosis based on fixed thresholds that ignore lift-off, scan direction, and prior magnetic history. This distinction should be stated explicitly in a review article because it shapes how the method ought to be validated \cite{Su2024MMMCivilEngineering}.

{ 
For deployment, the essential requirements are straightforward even if they are not easy to satisfy: repeatability across scan direction and speed, sensitivity studies with respect to stand-off and sensor orientation, and explicit treatment of nearby magnetic clutter. If the method output is binary, POD-style studies on blind defect or stress-concentration sets provide the cleanest bridge to accepted qualification practice \cite{ASTM_E2862_23,MIL_HDBK_1823_ASSIST}. If the output is a graded risk index, then the uncertainty in that index due to stand-off, baseline drift, and operator variability should be reported alongside the index itself. That requirement is not secondary. For passive self-field methods, it is the difference between a suggestive map and a defensible monitoring procedure.
}

%%%%%%%%%%%%%%%%%%

\subsection{ {Stress and fatigue damage monitoring}}
\label{sec:fatigue}

{ 
Fatigue is not a separate magnetic signal class; it is a monitoring workflow that can draw on passive self-fields, active magneto-mechanical response, and, at later stages, the onset of leakage fields associated with crack initiation and growth. In ferromagnetic steels, fatigue develops through microstructural evolution, residual-stress redistribution, and eventually crack nucleation and propagation. Magnetic signatures can therefore change before a macroscopic crack is easy to detect, but the mapping from magnetic response to fatigue state is strongly material and protocol dependent: steel grade, heat treatment, mean stress, load ratio, magnetization history, and sensor geometry all matter \cite{Bjorheim2022FatigueReview,jiles1995magnetomechanical,dapino2004villari,SantaAho2019BarkhausenReview}. This is why magnetic fatigue assessment is most defensible as a calibrated trending problem, not as a universal one-shot diagnostic.
}

{ 
The physical basis is magneto-mechanical coupling. Stress changes domain structure and effective permeability through magnetostriction (the Villari effect), while fatigue-driven microstructural evolution alters domain-wall pinning, hysteresis, remanence, and incremental permeability \cite{jiles1995magnetomechanical,dapino2004villari,Bao2012MagnetomechanicalFatigueSteel}. This is also why many magnetic fatigue indicators are loop-based rather than pointwise. If a magnetic field component $B$ is recorded together with strain $\varepsilon$ over fatigue cycle $N$, one useful cycle-resolved feature is the magnetic-loop area
}
\begin{equation}
A_{B\varepsilon}^{(N)}=\oint_{\text{cycle }N} B\,d\varepsilon,
\label{eq:fatigue_loop_area}
\end{equation}
{ 
together with a normalized trend indicator
}
\begin{equation}
D_N=\frac{A_{B\varepsilon}^{(N)}-A_{B\varepsilon}^{(0)}}{A_{B\varepsilon}^{(0)}}.
\label{eq:fatigue_damage_indicator}
\end{equation}
{ 
Such features are often more repeatable than raw field amplitudes because they reference each measurement to a controlled loading cycle and partially suppress slow baseline drift. The same logic applies when the loop is formed with force $F$ instead of strain.
}

{ 
Two measurement modes are common. Passive measurements track changes in residual or self-field under ambient bias and are convenient, but they inherit the ambiguity of passive magnetic methods discussed in Sec.~\ref{sec:MMM}. Active measurements impose a defined magnetization condition or synchronize the magnetic readout to the load cycle, which usually improves interpretability because the source field is better defined. For infrastructure monitoring, the practical lesson is simple: if the magnetic excitation is not controlled, the geometry and baseline must be controlled even more strictly.
}

\begin{figure}[!t]
\centering
\includegraphics[width=0.95\columnwidth]{  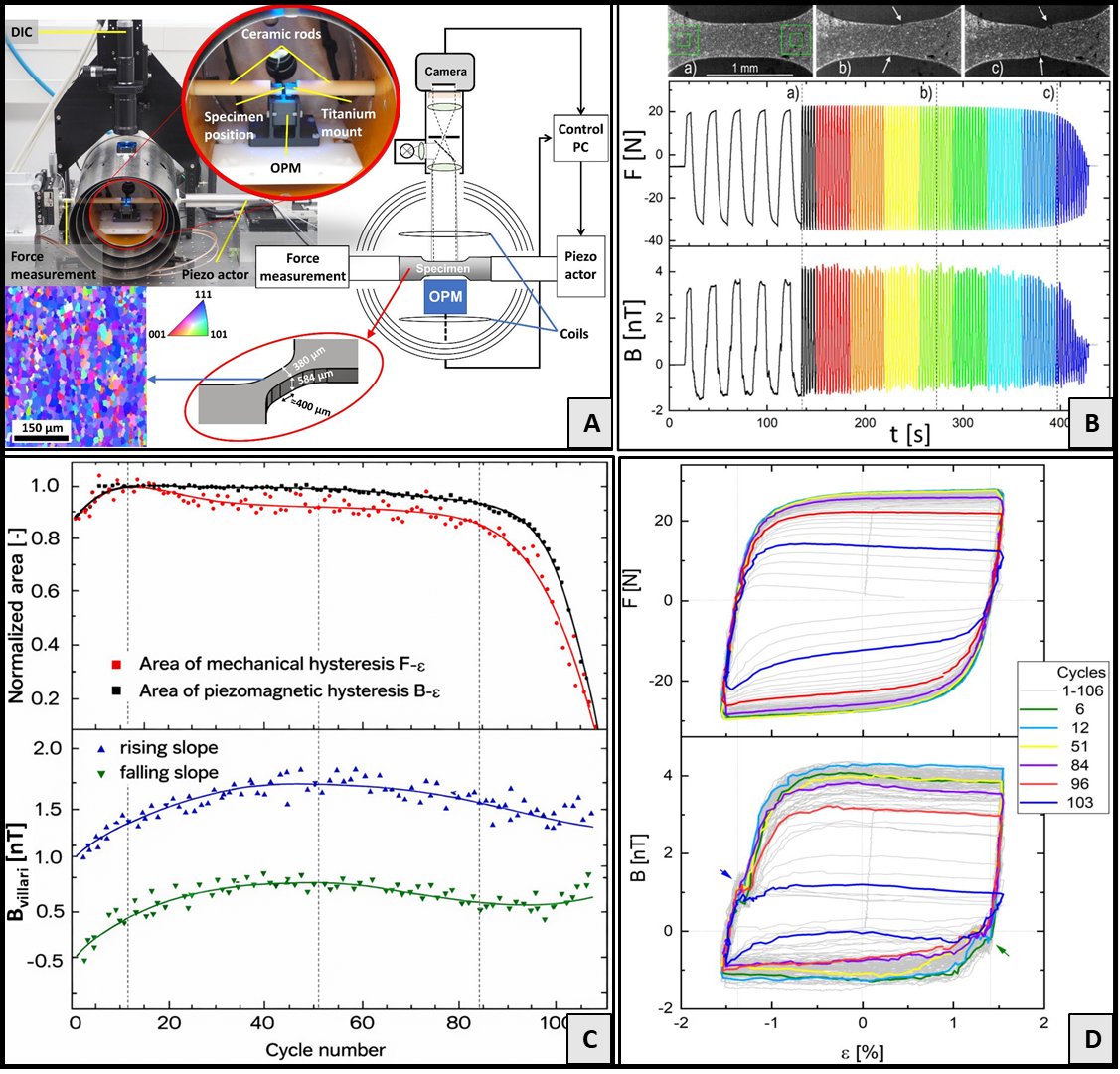}
\caption{ {Fatigue monitoring in steel using an optically pumped atomic magnetometer (OPM) as the magnetic receiver during cyclic loading. (A) Experimental setup showing the OPM position relative to the specimen, load frame, and optical strain measurement. (B) Example time traces of applied force $F$ (N) and measured magnetic signal $B$ (nT), with snapshots of crack evolution. (C) Cycle-dependent indicators extracted from magnetic hysteresis behavior, including normalized loop areas and bias-field trends. (D) Representative mechanical ($F$--$\varepsilon$) and magnetic ($B$--$\varepsilon$) hysteresis loops at selected cycle numbers \cite{koss2022optically}.}}
\label{fig:OPM_metal_fatigue_results}
\end{figure}

\paragraph{ {Load-synchronous magneto-mechanical measurements.}}

{ 
OPMs fit naturally into fatigue experiments because they preserve sensitivity in the low-frequency band where cyclic loading is usually monitored and can be arranged as gradiometers or reference-channel systems for background rejection. Figure~\ref{fig:OPM_metal_fatigue_results}(A) shows a fatigue-monitoring setup in which an OPM measures the magnetic flux density $B$ near a ferritic steel specimen during cyclic loading \cite{koss2022optically}. The important features are not only the sensor itself, but the controlled geometry and synchronized acquisition of force, strain, and magnetic field. With these observables aligned in time, the magnetic response can be examined cycle by cycle rather than as a drifting continuous trace.
}

{ 
Figure~\ref{fig:OPM_metal_fatigue_results}(B) shows time traces of $F(t)$ and $B(t)$ together with crack-evolution snapshots, illustrating why synchronization matters. Figure~\ref{fig:OPM_metal_fatigue_results}(C) shows cycle-dependent indicators extracted from the magnetic response, and Fig.~\ref{fig:OPM_metal_fatigue_results}(D) shows the mechanical and magnetic hysteresis loops from which those indicators are obtained. The scientific value of this result is clear: fatigue damage formation can perturb the cycle-resolved magnetic hysteresis before final fracture. The engineering interpretation should be more cautious. This is strongest today as a \emph{load-synchronous laboratory monitoring method under fixed geometry}, not yet as a fully mature free-running field monitor on arbitrary steel structures.
}

{ 
A portable magnetic fatigue instrument therefore needs more than a low noise floor. It needs a fixture or installation protocol that enforces repeatable sensor position and orientation, a reference channel or gradiometric baseline to reject common-mode drift, and a defined magnetization protocol when the method relies on active bias fields. Without those controls, lift-off variability and magnetic-history differences can dominate the indicators extracted from the loops. In other words, fatigue monitoring is a stability problem before it becomes a sensitivity problem.
}

\paragraph{ {NV mapping of localized fatigue-initiation sites.}}

{ 
NV sensors are best viewed as localized magnetic imagers for fatigue-related damage when close access is available. Their main contribution is not bulk stress measurement in steel, but near-surface mapping of stress concentration zones, early leakage-field signatures, heat-affected regions, or microstructural anomalies that can act as fatigue initiation sites. The same constraints recur: stable biasing, readout protocols that reject thermal drift through resonance splitting or differential tracking, and a scan plan that records stand-off and orientation. Unshielded damage imaging and tensor/gradient approaches (Fig.~\ref{fig:NVC_metal_crack_sensing_results}(I,II)) and grinding-burn or heat-affected-zone detection (Fig.~\ref{fig:NVC_metal_crack_sensing_results}(III)) are all relevant to fatigue for exactly this reason \cite{zhou2021imaging,newman2024tensor,vindolet5186260high}.
}

\begin{figure}[t]
\centering
\includegraphics[width=\linewidth]{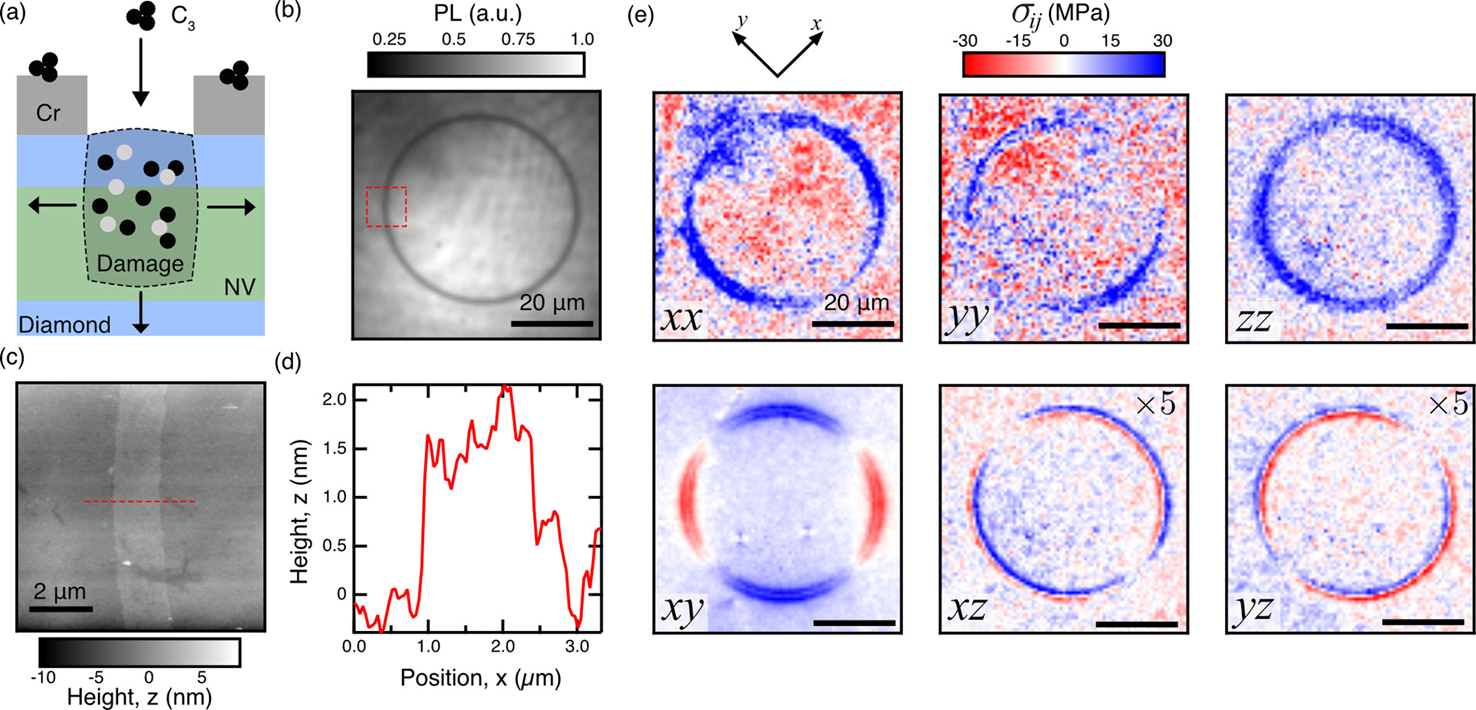}
\caption{NV-center stress-tensor imaging in diamond. (a) Conceptual geometry for generating and probing stress in the diamond host near an NV layer. (b) Photoluminescence (PL) image used to locate the region of interest (scale bar: 20~$\mu$m). (c) Atomic force microscopy (AFM) topography showing nanoscale surface deformation (height $z$ in nm; scale bar: 2~$\mu$m). (d) Line profile of the AFM height along the marked direction. (e) Reconstructed stress-tensor components $\sigma_{ij}$ (MPa), including normal components ($\sigma_{xx},\sigma_{yy},\sigma_{zz}$) and shear components ($\sigma_{xy},\sigma_{xz},\sigma_{yz}$), mapped over a 20~$\mu$m field of view using NV centers embedded in the diamond host \cite{broadway2019microscopic}.}
\label{fig:diamond_stress_NVC}
\end{figure}

Because fatigue discussions sometimes invoke the ``stress sensitivity'' of NV centers, the interpretation needs to remain explicit. Figure~\ref{fig:diamond_stress_NVC} does not show direct stress sensing in steel; it shows stress imaging \emph{inside the diamond itself} through spin--strain coupling of near-surface NV centers. In the work of Broadway \emph{et al.}, the ODMR response of an ensemble of in situ NV defects was used to reconstruct all six independent components of the local stress tensor with submicrometer spatial resolution and with reported sensitivities of order 1~MPa for shear components and 10~MPa for axial components \cite{broadway2019microscopic}. Figure~\ref{fig:diamond_stress_NVC}(a) summarizes the measurement concept, while panels~(b)--(d) show how the region of interest was identified and independently characterized through photoluminescence imaging and AFM topography before the full tensor field was reconstructed in panel~(e).

A key strength of the Broadway \emph{et al.} study is that it went well beyond a single proof-of-principle geometry. The same stress-imaging framework was applied to several distinct sources of stress inside or on diamond, including localized implantation damage, nanoindents, and scratches, and it also revealed unexpectedly large stress contributions from functional electronic structures fabricated on the diamond surface \cite{broadway2019microscopic}. In addition, the work showed sensitivity to deformations in materials brought into contact with the diamond. This makes the figure important as a demonstration of what NV centers can do as \emph{in situ} strain sensors within their host crystal: they can map complex, spatially varying stress fields with high spatial resolution and in a form directly relevant to diamond-device engineering.

The related work of Kehayias \emph{et al.} addressed a complementary regime that is particularly relevant for practical NV instrumentation \cite{kehayias2019imaging}. Instead of focusing on submicrometer mapping around localized nanoscale damage, they developed a micrometer-resolution, millimeter-field-of-view stress-imaging method for diamonds containing a thin surface layer of NV centers. Stress tensor elements were reconstructed over a two-dimensional field of view from ODMR spectra, and the resulting maps were used to study how stress inhomogeneity degrades NV magnetometry performance. In other words, the Kehayias \emph{et al.} work treated stress imaging not only as a materials-analysis tool but also as a diagnostic for sensor quality, crystal nonuniformity, and device optimization.

Across both OPM and NV platforms, the central limitation in fatigue applications is repeatability under realistic boundary conditions. Useful reports therefore do three things explicitly: they quantify uncertainty with respect to stand-off and orientation, they document the magnetization or loading protocol that defines the measured loop or trend, and they benchmark the magnetic indicators against established magnetic NDE tools or destructive post-mortem characterization on the same specimens \cite{SantaAho2019BarkhausenReview,NIST_MagneticSensingMetrology}.

Selected NV-center sensing workflows relevant to infrastructure inspection and monitoring are summarized in Table~\ref{tab:NV_applications}.

\begin{table}
\centering
\caption{ {Selected NV-center sensing workflows relevant to infrastructure inspection and monitoring. The table is organized by measurement role inside the workflow rather than by the sensor in isolation.}}
\label{tab:NV_applications}
\renewcommand{\arraystretch}{1.2}
{ 
\begin{tabular}{p{4cm} p{6.2cm} p{1.2cm}}
\hline
\textbf{Workflow / Signal Class} & \textbf{Outcome / Main Constraint} & \textbf{Representative Work} \\
\hline

Near-surface MFL-type field mapping on steel &
Vectorized stray-field maps with strong sensitivity to stand-off, motion, and magnetization geometry; best suited to accessible steel surfaces &
\cite{zhou2021imaging,newman2024tensor,villing2025implementing} \\

Portable or fiber-coupled point heads &
Reduced alignment burden and improved deployability; performance still depends on microwave delivery and thermal stability &
\cite{stuerner2021portable,Kubota2023WideTempNV} \\

Tensor / gradient background rejection &
Improved localization in cluttered environments; calibration depends on channel matching, baseline knowledge, and scan repeatability &
\cite{newman2024tensor} \\

Grinding-burn / microstructural anomaly mapping &
Sub-mm magnetic mapping of heat-affected zones or localized damage precursors; requires small stand-off and good surface access &
\cite{vindolet5186260high} \\

Stress imaging in diamond (mechanistic tool) &
Stress-tensor components mapped in the diamond host; informative for sensor physics and transduction concepts, but not direct steel-stress measurement &
\cite{kehayias2019imaging,broadway2019microscopic} \\

Battery eddy-current imaging and differential current sensing &
Lock-in imaging of defect-sensitive induced fields and robust current sensing on conductors or busbars; geometry and inversion assumptions remain central &
\cite{zhang2021battery,hatano2022high,Kubota2023WideTempNV} \\
\hline
\end{tabular}
}
\end{table}

\subsection{ {Asset-level synthesis: reinforced concrete structures}}
\label{sec:concrete}

{ 
Reinforced concrete is not a separate magnetic signal class; it is an asset family in which several signal classes coexist. Embedded rebar and post-tensioning tendons can be interrogated through passive self-fields, driven induction responses, and magnetization-based leakage fields, depending on access, cover depth, and the decision that must be made \cite{Eslamlou2023MagneticSensorsReview}. Corrosion and section loss can progress without obvious surface signs, especially in grouted ducts, anchor zones, and dense reinforcement layouts. Civil-infrastructure NDE is therefore usually layered: an areal screening method localizes suspect regions, followed by a higher-specificity method, selective opening, or both. Magnetic and electromagnetic approaches are attractive because low-frequency fields penetrate concrete cover and can be sensed from the surface without couplants.
}

\paragraph{ {Rebar corrosion and passive magnetic surveys.}}

{ 
Passive magnetic surveys, often described as self-magnetic flux leakage (SMFL) or MMM-like methods, measure the surface field under ambient geomagnetic bias and relate spatial features to rebar corrosion or stress concentration. Several studies report useful correlations between passive magnetic features and corrosion state, including links to non-uniform corrosion and cross-sectional loss \cite{Mosharafi2020RebarSMFL,Qiu2023SMFLUnevenness}. The limitation is the same as in other passive-self-field workflows: the measured field depends not only on corrosion but also on the prior magnetization of the steel, nearby ferromagnetic clutter, rebar spacing, and stand-off variations. In dense rebar mats, multiple bars can contribute comparable signatures, so vector or gradient measurements and a forward model are generally more defensible than fixed thresholds \cite{FrankowskiChady2022Magnetization,ISO24497_1}.
}

\paragraph{ {Driven induction through concrete cover.}}

{ 
Active electromagnetic methods reduce some of that ambiguity by imposing a known excitation and reading the secondary response. Eddy current testing (ECT) and related induction methods use a driven coil and record amplitude and phase changes that depend on geometry, conductivity, and permeability \cite{dealcantara2015ect}. In practical terms, deeper concrete cover pushes the useful excitation frequency downward, which reduces voltage pickup for conventional coils and makes phase stability harder in noisy environments. This is precisely the regime in which OPM receivers can be attractive, because they sense the magnetic response directly at low frequency rather than through $d\Phi/dt$ pickup. The usual cautions remain: a baseline over a known intact region is needed, multi-frequency operation is often preferable to a single-frequency inversion, and stand-off should be measured rather than assumed \cite{FrankowskiChady2023ACO,Eslamlou2023MagneticSensorsReview}.
}

\paragraph{ {Post-tensioning tendons and flux-based inspection.}}

Magnetization-based inspection is more mature for post-tensioning systems, where flux-based methods target strand fractures and section loss. In classical MFL, the tendon is magnetized toward saturation and sensors measure leakage fields produced by defects. In related ``main flux,'' ``return flux,'' or ``total flux leakage'' approaches, the instrument tracks changes in the overall flux path as a proxy for section loss, which can be more stable than leakage-only features in some tendon geometries \cite{karthik2019mfl,Kwahk2020TotalFluxLeakage}. FHWA development reports emphasize two persistent constraints: internal tendons are harder than external tendons because cover thickness and surrounding reinforcement introduce interfering magnetic paths, and field deployment requires fast, repeatable procedures with known magnetization conditions \cite{Lee2022FHWA_HRT23005,Ghorbanpoor2000FHWA_MFL}.

Quantum magnetometers fit into reinforced-concrete inspection mainly as receiver upgrades for these established signal classes. For low-frequency induction through concrete cover, OPM receivers can improve usable signal-to-noise where coil pickup becomes weak, but the system still needs lock-in detection, differential background rejection, and dynamic-range management in the presence of the geomagnetic field and moving steel. For accessible tendon inspection, NV sensors are attractive where small stand-off is feasible---for example on external tendons or exposed anchor regions---because the head can be compact and vector capable. Recent work has implemented NV-based sensing for MFL testing of prestressing steel and quantified sensitivity to displacement and lift-off, which is a useful step toward civil-infrastructure protocols \cite{villing2025implementing}. The key remaining gap is representative mockup validation in the presence of nearby rebars, ducts, anchors, and realistic concrete cover.

For both OPM and NV receivers, the near-term priority in reinforced concrete is comparative benchmarking on testbeds with known cover thickness, bar spacing, tendon layout, and controlled corrosion or section loss. Performance should be reported against conventional receivers in the same inspection geometry, with uncertainty budgets that explicitly include stand-off variability, magnetic-background drift, and inter-operator repeatability \cite{NIST_MagneticSensingMetrology,Lee2022FHWA_HRT23005}.

Selected OPM-relevant workflows, including established and prospective infrastructure applications, are summarized in Table~\ref{tab:OPM_applications}.

\begin{table}
\centering
\small
\caption{ {Selected OPM-relevant workflows for infrastructure inspection and monitoring. Rows marked ``prospective'' indicate application spaces where atomic receivers are plausible upgrades but still need dedicated validation on representative testbeds.}}
\label{tab:OPM_applications}
\renewcommand{\arraystretch}{1.25}
{ 
\begin{tabular}{p{4.5cm} p{6.5cm} p{3cm}}
\hline
\textbf{Workflow / Signal Class} & \textbf{Performance / Main Constraint} & \textbf{Representative Work} \\
\hline

Low-frequency EMI for concealed metal loss and CUI &
Phase-referenced amplitude/phase contrast through coatings or insulation in controlled scans; repeatability limited by lift-off, direct coil coupling, and layered geometry &
\cite{Bevington2018Steelwork,bevington2020inductive,maddox2022imaging} \\

Driven induction imaging with frequency tuning &
Low-frequency EMI with tunable depth sensitivity and direct magnetic readout; strongest where conventional coils become voltage-limited &
\cite{Deans2016ElectromagneticInductionRFAM,wickenbrock2016eddy} \\

Load-synchronous magneto-mechanical fatigue monitoring &
Cycle-resolved magnetic indicators during controlled cyclic loading; transferability depends on magnetization protocol and fixed geometry &
\cite{koss2022optically,Bao2012MagnetomechanicalFatigueSteel} \\

Gradiometric / unshielded operation &
Background suppression using multi-sensor layouts, compensation fields, and closed-loop control; essential for mobile scans in cluttered environments &
\cite{rushton2022unshielded,oelsner2022earthfield} \\

Battery induced-field diagnostics &
Non-contact induced-field mapping of cell-to-cell differences and anomalous internal states; scan fixture and field control are part of the method &
\cite{hu2020rapid} \\

Reinforced-concrete screening (prospective) &
Receiver upgrade for low-frequency induction through concrete cover and for controlled magnetic-gradient mapping; requires validation under rebar-interference conditions &
\cite{dealcantara2015ect,Eslamlou2023MagneticSensorsReview,Lee2022FHWA_HRT23005} \\

Portable / integrated OPM heads &
Microfabricated vapor cells, integrated optics, and compact coils enable fieldable heads, but thermal and field control remain instrument-level constraints &
\cite{oelsner2022earthfield,JimenezMartinez2017MicrofabricatedOPM} \\
\hline
\end{tabular}
}
\end{table}

\subsection{ {Operational-current fields and battery/electrical infrastructure}}
\label{sec:batteries}

{ 
Lithium-ion batteries and high-current electrical assets are especially useful benchmark problems because many early failure modes are fundamentally current-redistribution problems. Current crowding near tabs, partial connection faults, local shorts, and imbalance between parallel current paths can appear before they create a large change in total current or temperature. Magnetic sensing targets those modes directly because the external field is set by internal or operational current distribution through the Biot--Savart law. In this application family, it is helpful to keep two measurement modes separate from the start. The first is \emph{induced-field} or eddy-current imaging, in which an applied field interrogates the cell and the secondary response is measured. Conceptually, that belongs to the driven-induction class. The second is \emph{operando current-field sensing}, in which the magnetometer measures the field produced directly by the working current. That second mode belongs to the operational-current class proper \cite{hu2020rapid,zhang2021battery,hatano2022high}.
}

\paragraph{ {Induced-field and eddy-current battery diagnostics.}}

\begin{figure}[t]
\centering
\includegraphics[width=\linewidth]{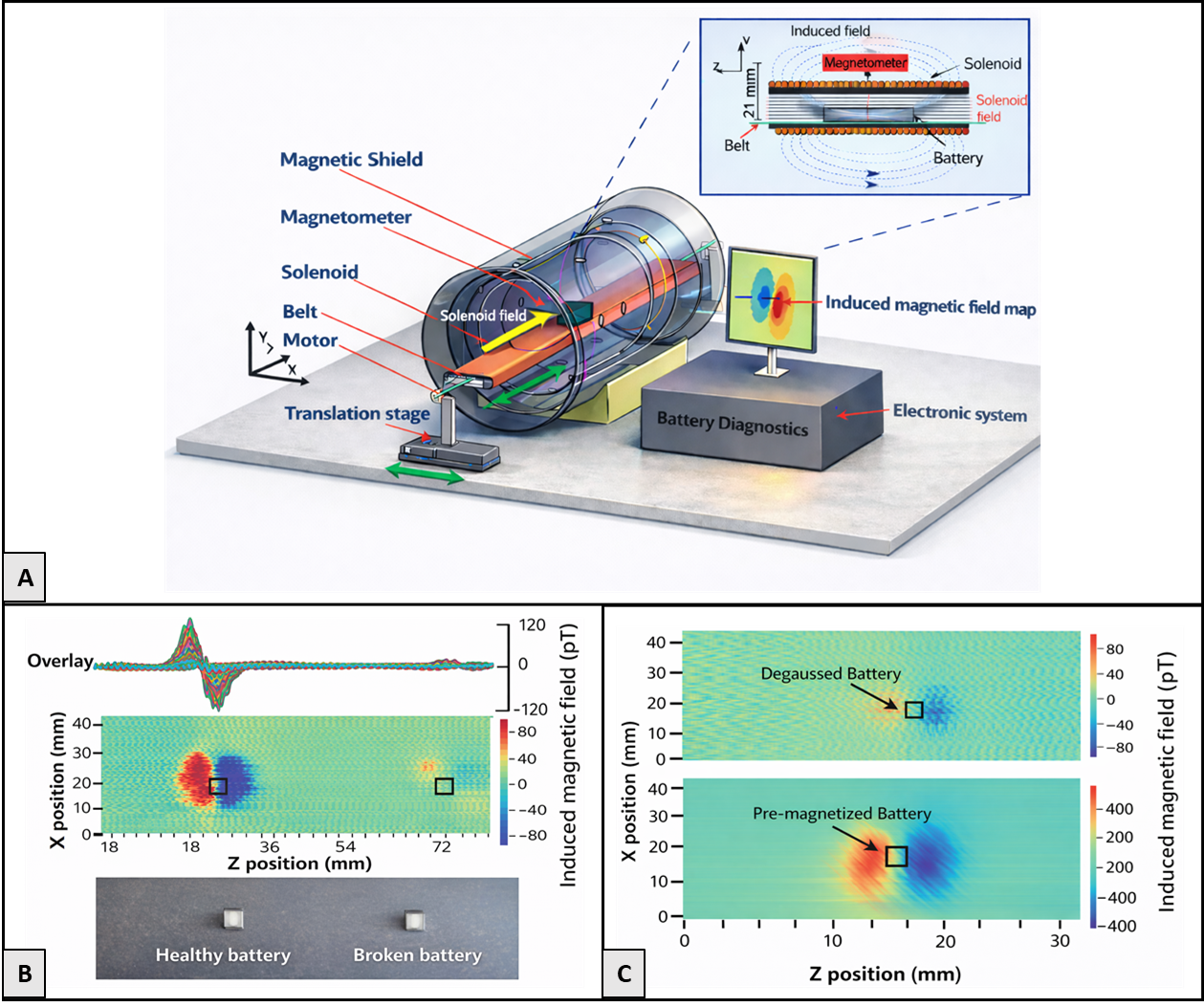}
\caption{Battery diagnostics using atomic magnetometry. (A) Experimental platform for induced-field imaging of miniature solid-state lithium-ion cells, using a long solenoid, magnetic shielding, a motorized translation stage, and an atomic magnetometer positioned in a low-primary-field region so that the measured signal is dominated by the battery-induced field. (B) Comparison of healthy and thermally damaged cells, including overlaid line scans, spatial maps of the induced magnetic field, and photographs showing that externally similar cells can be distinguished by their magnetic response. (C) Comparison of degaussed and pre-magnetized cells, demonstrating that magnetic history strongly affects contrast and supporting the interpretation that remanent or soft-ferromagnetic contributions are part of the measured signal \cite{hu2020rapid}.}
\label{fig:Battery_OPM_results}
\end{figure}

Hu \emph{et al.} used atomic magnetometry to diagnose miniature commercial solid-state batteries by measuring the field induced when the cell was exposed to a controlled background field inside a long solenoid \cite{hu2020rapid}. The key system idea, shown in Fig.~\ref{fig:Battery_OPM_results}(A), is an ``inside-out'' geometry: the battery is translated through the solenoid, while the atomic magnetometer is placed in a magnetically shielded region outside the effective primary field of the solenoid. In this arrangement, the sensor is largely insensitive to the directly applied field but remains sensitive to the smaller induced field produced by the battery itself. This allows the induced magnetic response of the cell to be mapped non-contactly, with the scan geometry defined by the conveyor-belt and translation-stage motion. In practical terms, the observable is still the induced-field difference map
\begin{equation}
\Delta B_{\mathrm{ind}}(\mathbf{r})=
B_{\mathrm{with\,cell}}(\mathbf{r})-
B_{\mathrm{reference}}(\mathbf{r}),
\label{eq:battery_induced_field}
\end{equation}
where the reference may be an empty scan, an intact cell, or the same cell after a controlled magnetic-history step.

Figure~\ref{fig:Battery_OPM_results}(B) shows the main diagnostic result of that work. Hu \emph{et al.} compared 40 cells in total, including 20 nominally healthy cells and 20 cells that had been thermally damaged by overheating above the specified operating temperature. The damaged cells consistently showed a different induced-field pattern and, in particular, a weaker induced response than the healthy cells, even though the cells were visually similar in ordinary photographs. The upper traces in Fig.~\ref{fig:Battery_OPM_results}(B) show the overlay of line scans, the middle panel shows the spatial field map, and the lower photograph emphasizes that the defect state is not evident from appearance alone. This is important for interpretation: the magnetic map is not merely imaging geometry, but is sensitive to the internal magnetic and materials state of the cell.

Figure~\ref{fig:Battery_OPM_results}(C) shows an especially useful control experiment. To test whether the measured contrast arose from intrinsic magnetic history rather than only from geometry, the authors compared degaussed healthy cells with healthy cells that were deliberately pre-magnetized. The pre-magnetized cells reproduced the stronger response seen in the nominally healthy population, whereas the degaussed cells showed a much weaker induced field, close to the behavior of the thermally damaged cells \cite{hu2020rapid}. Together with the paper's SQUID and compositional analysis, this supported the interpretation that remanent or soft-ferromagnetic contributions, likely associated with metallic nickel, play an important role in the induced-field contrast. For this reason, the figure is valuable not only as a battery-imaging demonstration, but also as a reminder that magnetic history can become an explicit state variable in battery diagnostics. That does not invalidate the method; it means that excitation history, degaussing or pre-magnetization protocol, and scan geometry must be documented if measurements are to be compared across cells or over time.

More broadly, this work shows why atomic magnetometers can be useful in battery metrology. The method does not directly image electrochemical state in the way that X-ray or magnetic-resonance methods do. Instead, it measures a weak field induced by the cell under controlled excitation and uses that field as a proxy for internal magnetic response, defect state, and manufacturing variability. This is most compelling when the geometry is repeatable, the magnetic history is controlled, and the induced field can be tied to an interpretable reference scan or classification protocol.

\begin{figure}[t]
\centering
\includegraphics[width=\linewidth]{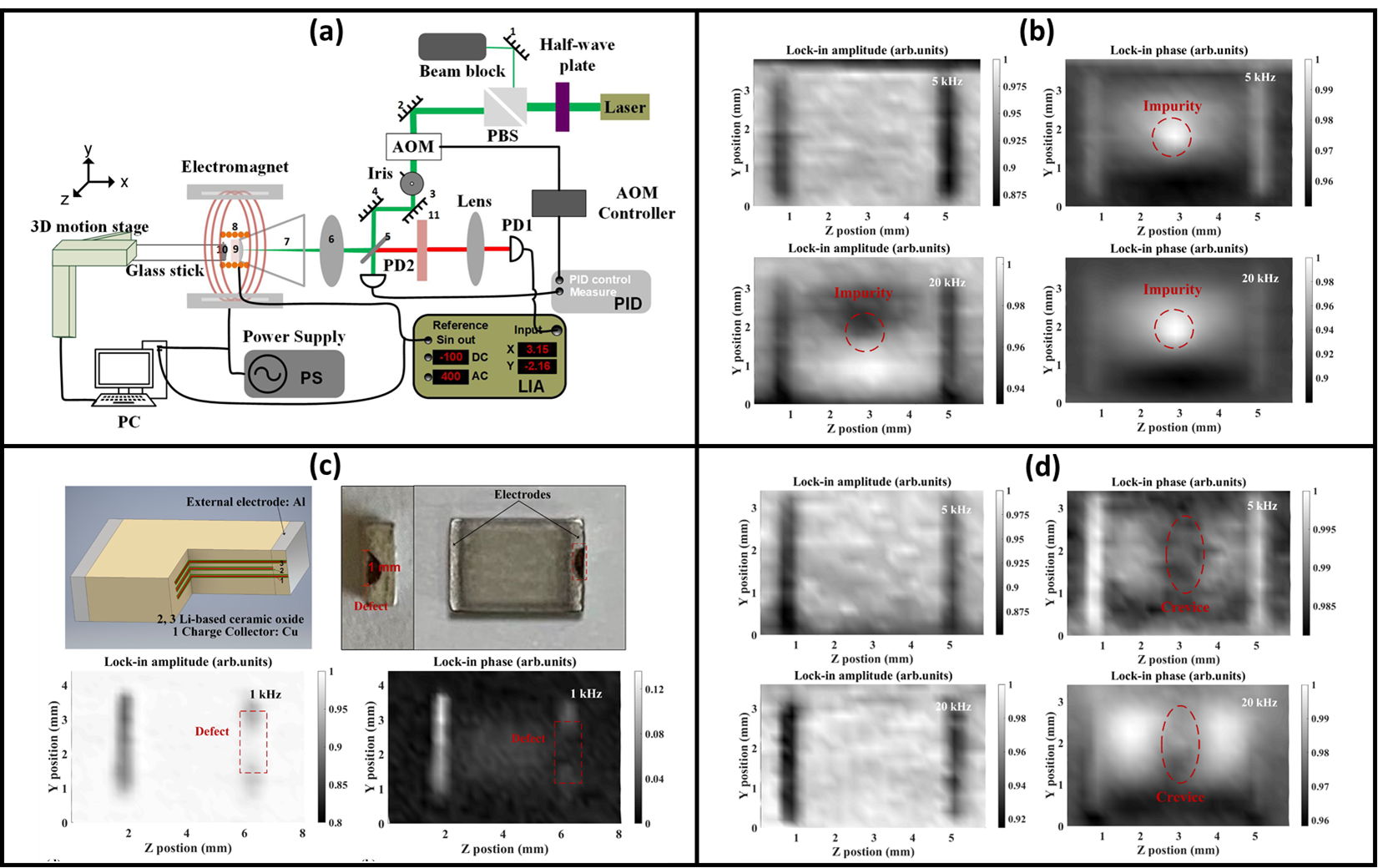}
\caption{NV-center eddy-current imaging of solid-state batteries. (a) Experimental setup for microwave-free NV-based eddy-current imaging with optical pumping, bias-field control, RF excitation, lock-in detection, and a scanning stage. (b) Battery structure and sample photograph highlighting an external defect, with corresponding lock-in amplitude and phase maps at 1 kHz. (c) Lock-in amplitude and phase images at 5 kHz and 20 kHz for a battery containing an internal impurity. (d) Lock-in amplitude and phase images at 5 kHz and 20 kHz for a battery containing an internal crevice, illustrating that hidden anomalies are most clearly expressed in the phase channel \cite{zhang2021battery}.}
\label{fig:Li_ion_battery_NVC_results}
\end{figure}

Zhang \emph{et al.} developed a complementary induced-field workflow based on NV centers in diamond, but with a notably different operating principle from standard ODMR-based NV magnetometry \cite{zhang2021battery}. Instead of microwave spectroscopy, they used microwave-free AC magnetometry based on a cross-relaxation feature between NV centers and substitutional nitrogen (P1) centers near 51.2~mT, with the experiment operated near a bias point of about 52.5~mT for improved sensitivity and reduced field dependence. The setup in Fig.~\ref{fig:Li_ion_battery_NVC_results}(a) combines continuous optical pumping, a controlled background field, an RF modulation coil, lock-in detection, and a motorized scan stage. In the experiment, the solid-state battery was scanned at a stand-off of about 0.1~mm from the diamond, so the measured lock-in amplitude and phase directly reflected the battery-induced secondary field rather than only the applied excitation.

Figure~\ref{fig:Li_ion_battery_NVC_results}(b) shows the baseline structural-imaging result. The sample was an all-ceramic multilayer solid-state battery with external aluminum electrodes, Li-based ceramic oxide inner electrodes and electrolyte, and a copper current collector. A deliberately introduced 1~mm defect on an external electrode was clearly visible in the lock-in maps. At 1~kHz, the phase image already captured the defective electrode geometry, while the amplitude and phase together gave a first estimate of how the battery redistributed the induced current field. This is the central methodological point of the figure: the measurement is not just an image, but a phase-referenced complex response.

Figures~\ref{fig:Li_ion_battery_NVC_results}(c) and \ref{fig:Li_ion_battery_NVC_results}(d) show why this distinction between amplitude and phase matters. To probe hidden internal anomalies, Zhang \emph{et al.} inserted two different artificial defects into otherwise similar batteries. In Fig.~\ref{fig:Li_ion_battery_NVC_results}(c), the anomaly is an impurity formed by an iron-containing brass cylinder of about 1~mm height and 1~mm diameter. In Fig.~\ref{fig:Li_ion_battery_NVC_results}(d), the anomaly is a crevice produced by cutting the battery and reattaching the two halves so that no obvious visual defect remained on the exterior. In both cases, the phase images at 5~kHz and 20~kHz reveal the internal anomaly more clearly than the amplitude images. The reason, discussed explicitly in the paper, is that the amplitude variations produced by these small internal defects were comparable to or smaller than slow background shifts, including temperature-related drift, whereas the phase channel preserved clearer defect-dependent contrast \cite{zhang2021battery}.

This work also makes the frequency dependence concrete. At lower excitation frequency, the measurement is more sensitive to larger-scale or deeper current paths, whereas at higher frequency the response becomes more weighted toward near-surface structure. In the Zhang \emph{et al.} data, the shift from 5~kHz to 20~kHz changes the visibility and apparent sharpness of the impurity and crevice signatures, showing that excitation frequency is not a secondary experimental setting but part of the sensing protocol itself. The paper further calibrated the induced response of the battery at 5~kHz, reporting a maximum magnetic field of about 0.04~mT and a phase shift of about 0.03~rad, and estimated a spatial resolution of about 360~\textmu m, limited mainly by the sensor--sample spacing \cite{zhang2021battery}. This is a useful benchmark because it ties the qualitative images in Fig.~\ref{fig:Li_ion_battery_NVC_results}(b)--(d) to concrete measurement performance.

\paragraph{ {Direct current monitoring and current-density inversion.}}

\begin{figure}[t]
\centering
\includegraphics[width=0.8\linewidth]{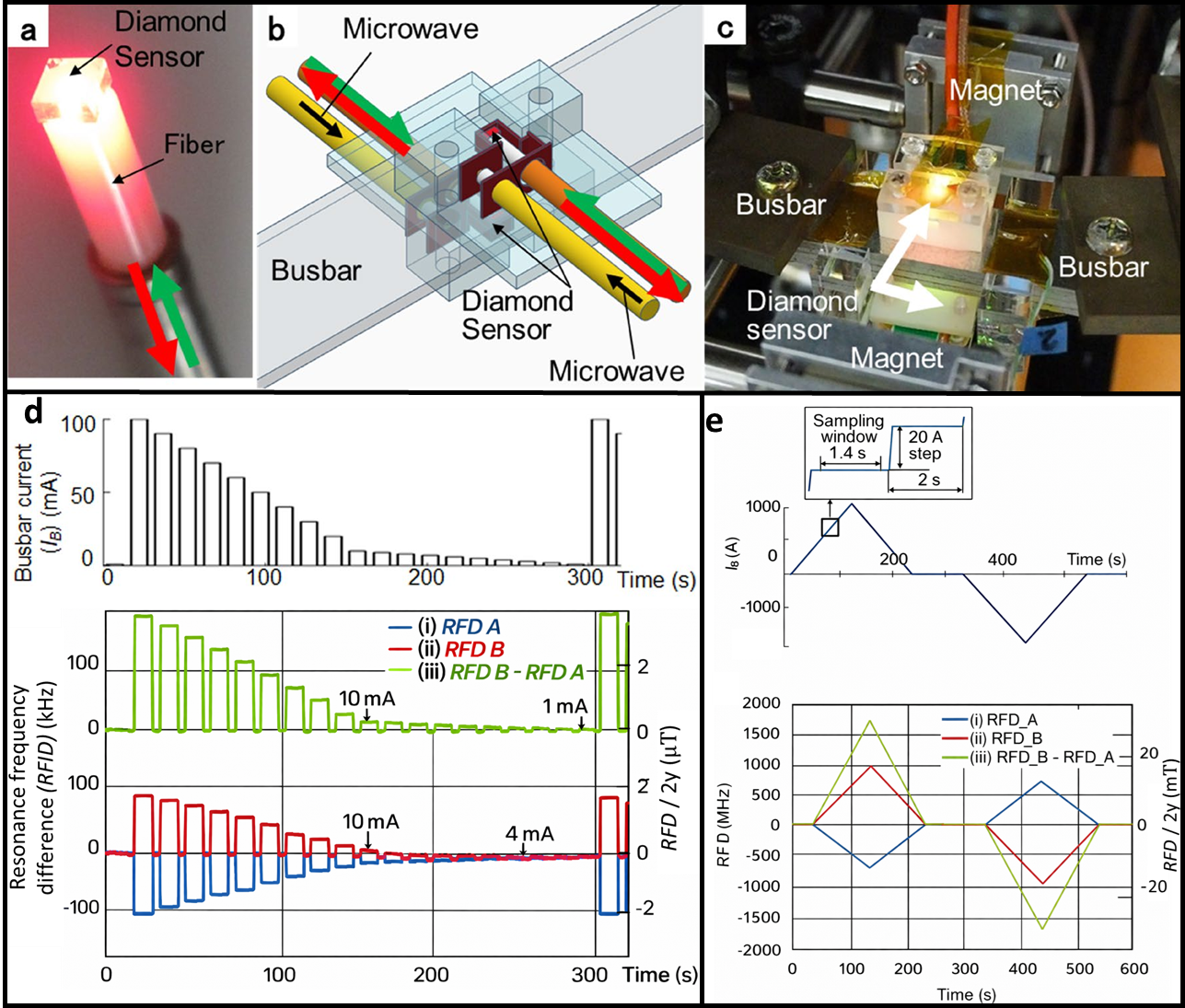}
\caption{ {NV-center magnetometry for high-precision current monitoring in battery-relevant conductors and busbars. (a) Fiber-coupled NV-diamond sensor head, in which a small diamond element is attached to the end of an optical fiber for compact local magnetic-field detection. (b) Differential sensing concept using two NV sensors on opposite sides of a current-carrying busbar so that the busbar field adds while common-mode environmental noise largely cancels. (c) Photograph of the realized busbar sensor assembly, including the two diamond heads and permanent magnets used to set the bias field. (d) Time-resolved resonance-frequency-difference (RFD) response during small-current step tests, showing differential detection of low busbar currents in a noisy environment. (e) RFD response during large-amplitude dynamic current excitation, illustrating tracking of current magnitude, sign, and waveform over a wide dynamic range \cite{hatano2022high}.}}
\label{fig:EV_Battery_NVC_results}
\end{figure}

{ 
Direct current monitoring on conductors such as tabs, busbars, and module interconnects is one of the clearest near-term use cases for quantum magnetometry because the field source is well defined and the electrical ground truth can be measured independently. The work of Hatano \emph{et al.} is important in this context because it did not treat the NV sensor as a laboratory magnetometer in isolation; it treated it as a current-monitoring subsystem intended for electric-vehicle battery modules \cite{hatano2022high}. Figure~\ref{fig:EV_Battery_NVC_results}(a) shows the compact fiber-coupled sensor head, where a small NV-diamond element was attached to the end of a multimode fiber and combined with a microwave antenna. Figure~\ref{fig:EV_Battery_NVC_results}(b) then shows the central systems idea: two such sensor heads are placed on opposite sides of a copper busbar, so that the magnetic field generated by the busbar current has opposite sign at the two sensors, while a large part of the external magnetic noise is shared by both sensors. Figure~\ref{fig:EV_Battery_NVC_results}(c) shows the realized experimental assembly, including the busbar, the two diamond heads, and the permanent magnets used to define the static bias field for ODMR operation.
}

{ 
The differential observable can be written as
\begin{equation}
B_{\mathrm{diff}}(t)=\big[B_1(t)-B_2(t)\big]\cdot\hat{\mathbf{u}}
\approx G_I\,I(t)+\epsilon_{\mathrm{cm}}(t),
\label{eq:busbar_diff}
\end{equation}
where $G_I$ is a calibration factor set by sensor placement and axis alignment, and $\epsilon_{\mathrm{cm}}$ is the residual common-mode term after subtraction. In the Hatano \emph{et al.} implementation, the measured quantity was not raw field itself but the resonance-frequency difference (RFD) extracted from the ODMR spectra of each NV sensor. The two single-sensor channels, labeled RFD$_A$ and RFD$_B$ in Fig.~\ref{fig:EV_Battery_NVC_results}(d,e), move in opposite directions under busbar current, while the differential combination RFD$_B-$RFD$_A$ isolates the busbar signal and suppresses common-mode disturbance \cite{hatano2022high}. This is exactly the form of receiver-level background rejection that makes the application attractive: the system is designed so that the useful signal is antisymmetric, while much of the environmental noise is symmetric.
}

{ 
Figure~\ref{fig:EV_Battery_NVC_results}(d) shows why this geometry matters for sensitivity. In that experiment, a train of small busbar-current steps was applied, decreasing from the 100~mA scale toward the low-mA regime over time. The single-sensor responses remain visible, but the differential channel is the most informative because it rejects shared background fluctuations and preserves the expected step sequence \cite{hatano2022high}. This is the key result for practical EV sensing: the sensor must detect small current changes even when mounted in an electrically noisy environment. Hatano \emph{et al.} identified 10~mA as the target current resolution relevant to state-of-charge estimation and showed that differential detection was sufficient to recover this scale in time-domain measurements under realistic noise conditions.
}

{ 
Figure~\ref{fig:EV_Battery_NVC_results}(e) addresses the opposite requirement: wide dynamic range. Here the input current was driven over a large bipolar waveform, ultimately reaching the $\pm 1000$~A class in stepped tests from an external current source \cite{hatano2022high}. The point is not only that the sensor sees large current, but that it remains locked while doing so. To achieve this, the authors used mixed analog--digital control of the microwave-generator frequency so that the NV resonance could be tracked over more than 1~GHz of frequency excursion. In the resulting data, the differential RFD channel followed the applied waveform correctly across the full current sweep, and the paper reported linearity within about $\pm 0.3\%$ over the high-current operating range together with magnetic-noise fluctuation corresponding to about 10~mA/$\sqrt{\mathrm{Hz}}$ in the differential output \cite{hatano2022high}. In other words, the same architecture that resolved low-mA changes also remained functional when the busbar current was increased by five orders of magnitude.
}

{ 
A second practical result from the same work is that the system was tested under conditions relevant to EV battery monitoring rather than only with benchtop current sources. Battery-module current was tracked up to about 130~A under driving-pattern-relevant conditions, and the authors argued that the resulting current accuracy was sufficient to reduce the state-of-charge margin required in present EV systems \cite{hatano2022high}. The setup therefore bridges laboratory quantum sensing and application-specific current metrology in a way that is uncommon in the literature.
}

{ 
Subsequent work by Kubota \emph{et al.} strengthened the deployment case by focusing on thermal robustness of the sensor head itself \cite{Kubota2023WideTempNV}. They developed a compact NV-diamond current-sensor head with dimensions on the order of $1\times1\times0.5~\mathrm{cm}^3$, confirmed 10~mA accuracy with 500~Hz bandwidth, and demonstrated stable operation over a very wide temperature range, extending well beyond ordinary automotive conditions. That result is important because it shifts the discussion from proof-of-principle current sensing to package-level survivability and performance stability under realistic thermal environments.
}

{ 
When the goal is not a scalar current reading but a current-density image, the inverse problem becomes central. For a two-dimensional current sheet at stand-off $h$, one commonly used Fourier-domain relation is
\begin{equation}
\tilde{B}_z(\mathbf{k},h)
=
\frac{\mu_0}{2}\,e^{-kh}\,
\frac{i}{k}\left(k_x \tilde{J}_y-k_y \tilde{J}_x\right),
\qquad
k=\sqrt{k_x^2+k_y^2},
\label{eq:current_inversion_fourier}
\end{equation}
which makes the ill-conditioning explicit: large spatial frequencies are exponentially suppressed by stand-off \cite{Roth1989MagnetometerImaging,Midha2024CurrentDensityRecon}. In practice, reconstructions therefore depend strongly on geometry constraints, regularization, and accurate knowledge of the measurement plane. This is why direct differential current sensing on conductors, as in Fig.~\ref{fig:EV_Battery_NVC_results}(a)--(e), is currently a more deployment-ready target than unconstrained current-density imaging in large battery modules.
}

For clarity, the main observables used across OPM and NV workflows and their corresponding engineering interpretations are summarized in Table~\ref{tab:engineering_mapping}.

\begin{table}
\centering
\small
\caption{ {Engineering interpretation of quantum-magnetometer observables used in infrastructure and energy diagnostics. Each observable must still be paired with a geometry model (sensor position and orientation, stand-off, and any applied excitation) and an uncertainty estimate.}}
\label{tab:engineering_mapping}
\renewcommand{\arraystretch}{1.25}
{ 
\begin{tabular}{p{4.2cm} p{4.3cm} p{5.0cm}}
\hline
\textbf{Measured Observable} & \textbf{Primary Quantity} & \textbf{Engineering Use} \\
\hline

OPM / AVM lock-in amplitude and phase at excitation frequency &
Secondary-field response from driven eddy currents (complex transfer function) &
Thickness and conductivity contrast in CUI, hidden metal loss, and low-frequency induction-based screening \\

Induced-field difference map $\Delta B_{\mathrm{ind}}(\mathbf{r})$ &
Change in effective magnetic susceptibility or induced-current distribution relative to a reference state &
Battery state or defect contrast; localization of anomalous regions; comparison between nominally identical cells \\

Vector field map $\mathbf{B}(\mathbf{r})$ &
Stray field from magnetization distortions or operational currents &
Defect localization in MFL-type inspection; current-path mapping in batteries, busbars, and conductors \\

NV ODMR splitting $(f_{+}-f_{-})$ &
Magnetic-field component along an NV axis ($B_{\parallel}$) with partial rejection of common-mode drift &
Local magnetic signatures from defects or currents; background-reduced magnetometry in compact heads \\

Differential two-sensor output $B_{\mathrm{diff}}$ &
Geometry-calibrated current proxy from opposing sensor channels &
Robust current monitoring on busbars or interconnects with common-mode noise suppression \\

NV ODMR average $(f_{+}+f_{-})/2$ after magnetic contribution removal &
Shift of $D$ dominated by temperature or strain in the diamond &
Sensor-head diagnostics and drift compensation; not a direct infrastructure observable without transduction calibration \\

Gradient or tensor estimate &
Spatial derivatives of $\mathbf{B}$ &
Background suppression and sharper localization of small anomalies, provided sensor baseline and calibration are known \\

Cycle-resolved loop area $A_{B\varepsilon}^{(N)}$ or related trend indicators &
Load-synchronous magneto-mechanical response under controlled boundary conditions &
Fatigue trending and damage evolution studies in ferromagnetic components when geometry and loading are controlled \\

\hline
\end{tabular}
}
\end{table}

%%%%%%%%%%%%%
\section{ {Deployment, qualification, and outlook}}
\label{sec:outlook}

{ 
Quantum magnetometers matter most when receiver limits dominate the measurement chain: low-frequency signals that are hard for inductive pickups, measurements through coatings or insulation where stand-off cannot be made vanishingly small, and environments in which time-varying magnetic backgrounds compete directly with the task signal. Even in those cases, sensitivity is only useful when the measurement is repeatable. Repeatability comes from controlled geometry, background rejection, and a documented path from a measured observable---a field component, a gradient, or a complex lock-in response---to an engineering metric with uncertainty \cite{Fu2020ChallengingEnvironments,NIST_MagneticSensingMetrology}.
}

{ 
The applications closest to robust deployment are those with relatively simple source geometry and accessible ground truth. Differential current monitoring on busbars and battery interconnects is the clearest example. The field source is well defined, sensor placement can be constrained, and the reference current can be measured with calibrated instrumentation. The differential NV configuration in Fig.~\ref{fig:EV_Battery_NVC_results}(b--e) and the wide-temperature follow-on work for EV conditions show the systems features that make this promising: common-mode rejection by geometry, controlled biasing, and resonance tracking that remains locked over a wide current range \cite{hatano2022high,Kubota2023WideTempNV}. Induced-field battery diagnostics are also near term, but mainly in controlled environments where the excitation field, shielding or compensation, and scan geometry can be kept stable \cite{hu2020rapid,zhang2021battery}. For structural assets, low-frequency induction screening and accessible MFL-style mapping on tendons or exposed steel are the most plausible early deployment paths \cite{bevington2020inductive,Lee2022FHWA_HRT23005,villing2025implementing}.
}

{ 
The harder cases remain the ones in which the source field itself is poorly controlled. Passive self-field methods for corrosion or stress concentration are attractive because they simplify the hardware, but that convenience shifts the burden to interpretation and repeatability. Magnetic history, lift-off, geomagnetic orientation, and nearby ferromagnetic clutter all become first-order variables, which limits transferability across assets and across measurement days \cite{Feng2022MFLReview,ISO24497_1,ISO24497_2}. These methods may still be valuable for controlled trending on the same asset, but they are presently the least mature path to widely accepted qualification.
}

{ 
Qualification should therefore be treated as a first-class output of quantum-infrastructure sensing research, not as a future afterthought. A credible field instrument needs more than a sensitivity number. It needs: (i) a stated task-band noise and dynamic-range specification under realistic background fields, (ii) explicit metrology for stand-off, orientation, and scan position, (iii) calibration artifacts representative of the intended asset and defect family, (iv) repeatability studies across days and operators, and (v) blind validation on representative defect sets. Where the output is binary, probability-of-detection (POD) methodology provides the most direct bridge to accepted NDE practice; where the output is continuous, sizing error or uncertainty bands should be reported alongside the estimate itself \cite{ASTM_E2862_23,MIL_HDBK_1823_ASSIST,NIST_MagneticSensingMetrology}.
}

{ 
The most promising instrument concept is therefore not ``sensor only'' but ``sensor head plus geometry metrology plus referencing.'' In practical terms, that means integrating the magnetic receiver with stand-off control or ranging, encoder or pose metadata, a reference or gradiometric channel, and closed-loop readout that remains in range while the sensor moves. Arrays will become increasingly important because they provide spatial filtering and common-mode rejection in hardware, but array calibration---gain matching, axis alignment, and drift tracking---must be part of the instrument design rather than post-processing wishful thinking.
}

{ 
Data-driven methods will help, but mainly after the physical chain is explicit. The most useful near-term roles for machine learning are fast resonance estimation, nuisance-parameter estimation, and physics-constrained inversion. Current-density reconstruction from wide-field quantum magnetic maps is a good example: Bayesian or learning-assisted methods can stabilize an ill-posed inverse problem, but they do not remove the exponential loss of high-spatial-frequency information with stand-off \cite{Midha2024CurrentDensityRecon}. The same principle applies in MFL and induction imaging. Learned inference becomes valuable only when geometry, calibration, and nuisance-variable coverage are already built into the dataset and validation plan.
}

{ 
The platform-specific outlook is now fairly clear. OPMs are most likely to mature first as low-frequency, phase-referenced receivers for driven induction and for fixed-geometry, load-synchronous magneto-mechanical monitoring. NV sensors are most likely to mature first as compact solid-state heads for near-surface vector or gradient mapping and for differential sensing of operational currents. Neither platform changes the stand-off decay of leakage fields, the ill-posedness of current inversion, or the magnetization dependence of ferromagnetic inspection. The decisive step toward accepted infrastructure use will therefore be robust instrument packaging, explicit geometry control, and qualification procedures that resemble those used for other NDE systems rather than demonstrations that emphasize sensitivity alone.
}

\section{ {Conclusion}}
\label{sec:conclusion}

{ 
Quantum magnetometers are best understood as receiver technologies that can strengthen established magnetic inspection signals rather than replace established workflows. In infrastructure and energy monitoring, the most relevant use cases fall into four signal classes: driven induction responses used to infer thickness or conductivity, leakage fields from magnetization distortions used in MFL-like inspection, passive self-fields linked to stress or corrosion state, and operational-current fields used in batteries and power assets.
}

{ 
The comparative roles of the two room-temperature platforms are now distinct enough to state clearly. OPM receivers are presently strongest for low-frequency, phase-referenced induction measurements and for load-synchronous magneto-mechanical measurements in which conventional coils become voltage-limited or phase unstable at low frequency. NV receivers are presently strongest for near-surface stray-field mapping, vector or gradient measurements at small stand-off, and differential current sensing on conductors and busbars when compact solid-state packaging and wide dynamic range matter most. Neither platform removes the lift-off, magnetization, or inversion limits set by the asset physics itself.
}

{ 
Across all signal classes, adoption depends more on repeatable geometry control, background rejection, and calibrated interpretation than on laboratory noise floors. The highest-value future work will therefore be the work that reports stand-off and orientation uncertainty, integrates referencing and geometry metrology into the scan head, validates against representative artifacts, and benchmarks quantum receivers against established receivers in the same inspection geometry. That is the path by which quantum magnetometry is most likely to move from promising demonstrations to accepted infrastructure practice.
}

% ---------------------------
% Declarations
% ---------------------------

\section*{CRediT authorship contribution statement}

Muhammad Mahmudul Hasan: Writing -- original draft, Investigation, Formal analysis, Visualization. 
Ingrid Torres: Writing -- review \& editing, Investigation, Visualization. 
Alex Krasnok: Conceptualization, Supervision, Funding acquisition, Project administration, Writing -- review \& editing.

\section*{Declaration of competing interest}

The authors declare that they have no known competing financial interests or personal relationships that could have appeared to influence the work reported in this paper.

\section*{Acknowledgments}

The authors acknowledge financial support from the U.S. Department of Energy (DoE) and the U.S. Air Force Office of Scientific Research (AFOSR).

\section*{Data availability}

No new experimental data were generated.

\bibliographystyle{unsrt}
\bibliography{references}

\end{document}